\documentclass[12pt]{article}

\usepackage{graphics,epsfig}
\font\mybbb=msbm10 at 8pt

\def\bbb#1{\hbox{\mybbb#1}}
\def\bb#1{\hbox{\mybb#1}}
\def\pRe{\bbb{R}}

\def\dalemb#1#2{{\vbox{\hrule height .#2pt
        \hbox{\vrule width.#2pt height#1pt \kern#1pt
                \vrule width.#2pt}
        \hrule height.#2pt}}}

\def\half{{\textstyle{1\over2}}}
\let\a=\alpha \let\b=\beta \let\g=\gamma \let\d=\delta \let\e=\epsilon
\let\z=\zeta  \let\th=\theta  \let\k=\kappa
\let\l=\lambda \let\m=\mu  \let\x=\xi \let\p=\pi %\let\r=\rho
\let\s=\sigma \let\t=\tau    
\let\vp=\varphi \let\vep=\varepsilon
\let\w=\omega       \let\D=\Delta \let\Th=\Theta \let\L=\Lambda
\let\X=\Xi \let\P=\Pi \let\S=\Sigma  \let\Y=\Psi
\let\C=\Chi \let\W=\Omega
\let\la=\label \let\ci=\cite 
  
\def\nn{\nonumber} \def\bd{\begin{document}} \def\ed{\end{document}}
\def\ds{\documentstyle} \let\fr=\frac \let\bl=\bigl \let\br=\bigr
\let\Br=\Bigr \let\Bl=\Bigl
\let\bm=\bibitem
\let\na=\nabla
\def\tU{{\widetilde U}}
\let\pa=\partial \let\ov=\overline
\def\ie{{\it i.e.\ }}
\newcommand{\be}{\begin{equation}}
\newcommand{\ee}{\end{equation}}
\def\ba{\begin{array}}
\def\ea{\end{array}}
\def\ft#1#2{{\textstyle{{\scriptstyle #1}\over {\scriptstyle #2}}}}
\def\fft#1#2{{#1 \over #2}}
\def\F#1#2{{ F_{#1}^{(#2)} }}
\def\cF#1#2{{ {\cal F}_{#1}^{(#2)} }}

\def\={\, =\, }
\def\+{\, +\, }
\def\-{\, -\, }

\def\R{{\bf R}}
\def\sst#1{{\scriptscriptstyle #1}}
\def\oneone{\rlap 1\mkern4mu{\rm l}}
\def\e7{E_{7(+7)}}
\def\td{\tilde}
\def\wtd{\widetilde}
\def\im{{\rm i}}
\newcommand{\ho}[1]{$\, ^{#1}$}
\newcommand{\hoch}[1]{$\, ^{#1}$}
\newcommand{\bea}{\begin{eqnarray}}
\newcommand{\eea}{\end{eqnarray}}
\newcommand{\ra}{\rightarrow}
\newcommand{\lra}{\longrightarrow}
\newcommand{\Lra}{\Leftrightarrow}
\newcommand{\ap}{\alpha^\prime}
\newcommand{\bp}{\tilde \beta^\prime}
\newcommand{\cB}{{\cal B}}
\newcommand{\cO}{{\cal O}}
\newcommand{\vecx}{\vec{x}}
\newcommand{\vecy}{\vec{y}}
\newcommand{\vecp}{\vec{p}}
\newcommand{\vecq}{\vec{q}}
\newcommand{\tr}{{\rm tr} }
\newcommand{\Tr}{{\rm Tr} }

\newcommand{\cL}{{\cal L}}
\newcommand{\cA}{{\cal A}}
\newcommand{\cD}{{\cal D}}
\def\sst#1{{\scriptscriptstyle #1}}
\def\0{{\sst{(0)}}}
\def\1{{\sst{(1)}}}
\def\2{{\sst{(2)}}}
\def\3{{\sst{(3)}}}
\def\4{{\sst{(4)}}}
\def\5{{\sst{(5)}}}
\def\6{{\sst{(6)}}}
\def\7{{\sst{(7)}}}
\def\8{{\sst{(8)}}}
\def\ve{\varepsilon}
\def\vf{\varphi}
\def\F{\Phi}
\def\wg{\wedge}

\def \nn {\nonumber}
\def \rk  {m}
\def \L {{\Lambda}}
\def \ka  { {\kappa }}
\def \S {{ \call S}}

\def\up{\uparrow}
\def\down{\downarrow}

\def \foot {\footnote}
\def \bi{\bibitem}

\def \tr {{\rm tr}}
\def \ha {{1 \over 2}}
\def \td {\tilde}
\def \ci{\cite}
\def \N {{\mathcal N}}
\def \ww {\Omega}
\def \const {{\rm const}}
\def \ss {\sum_{i=1}^3 }
\def \t {\tau}
\def\S{{\mathcal S} }
\def \XX {{\rm X}}

\def \lra {\leftrightarrow}
\def \vom {{\bar \omega}}
\def \E {{\mathcal  E}} \def \J {{\mathcal  J}}
\def \YY {{\rm Y}}

\def \d {\del}
\def \rJ {{J}}
\def \sms {sigma models\ }
\def \sm {sigma model\ }
\def \L {\Lambda}
\def \gl {\ell}
\def \tr {{\rm tr\ }}
\def\z{\zeta}
\def\zi{\zeta_1}
\def\zii{\zeta_2}
\def\K{\mbox{K}}
\def\eE{\mbox{E}}   \def \vt {\vartheta}
\def \vr {\varrho}
\def \wup {w}

\def\dg{\dagger}
\def\a{\alpha}
\def\b{\beta}
\def\e{\varepsilon}
\def\p{\phi}
\def\ap{\alpha^\prime}
\def\I{{\cal I}}

\def\R{{\bf R}}
\def\Z{{\bf Z}}
\def\C{{\bf C}}
\def\P{{\bf P}}
\def\xb{{\bar X}}
\def\Tr{{\rm  Tr}}
\def\tr{{\rm  tr}}

\def \del{\partial}
\def \a {\alpha}
\def \aa {{\a'}}
\def\g{\gamma}
\def\s{\sigma}
\def\z{\zeta}
\def\zi{\zeta_1}
\def\zii{\zeta_2}
\def\ov{\over}

\def\I{{\cal I}}
\def\J{{\mathcal J}}
\def \ok {{1\ov \k}}
\def\LL{{\mathcal L }}
\def \jL {{J}}
\def \om {\omega}
\def \cL {{\mathcal L}} \def \cH {{\mathcal H}}
\def\E{{\mathcal E}}
\def\w{\omega}
\def\b{\beta}
\def\l{\lambda}
\def\eps{\epsilon}
\def\vep{\varepsilon}
\def \De {{\mathcal D}}

\def  \Jt {  {J}_{\rm tot}    }

\def \k {\kappa}
\def\foot{\footnote}
\def \four{{\textstyle {1\ov 4}}}
 \def \third { \textstyle {1\ov 3
}}
\def\det{\hbox{det}}
\def \ci {\cite}

\def \foot {\footnote}
\def \bi{\bibitem}

\def \tr {{\rm tr}}
\def \ha {{1 \over 2}}
\def \tid {\tilde}
\def \vv {{\rm v}}
\def \tl {{\tilde \l}}
\def \XX {{\rm X}}
\def \ta {{\tilde \a}}
\def \fo { {1\ov 4}}
\def \ep {\epsilon}
\def \inti {{\int^{2\pi}_0 {d \sigma \ov 2 \pi}}}

\def \d {\partial}
\def \K {{\rm S}}
\def \el {\ell}
\def \Tr {{\rm Tr}}
\def \P {\Phi}
\def \l  {\lambda}
\def \tl {{\tilde \l}}
\def \bl {{\tilde \l}}
\def \const {{\rm const}}
\def \V {v}

\def \bv {v^*}
\def \vv {{\rm v}}
\def \LL {{\mathcal L}}
\newcommand{\PV}[1]{P_{\!\!_{V_{#1}}}}

\def \bL {\ell}
\def \M {{\mathcal M}}
\def \N {{\mathcal N}}
\def \S {{\rm S}}
\def \vn {\vec n}
\def \tl {\td \l}
\def \td {\tilde}
\def \Prod {\Pi}
\def \O {{\mathcal O}}
\def \Q {{\rm  Q}}
\def \D {\Delta}
\def \N {{\mathcal N}}
\def\tN{{\tilde N}}

\def \m {\mu}
\def \vs {\vec \s}
\def \ie {i.e.}

\def \cD {{\cal D}}

\def  \le  {\l_{\rm eff}}

\def \rS {{\rm S}}
\def\as{{\a}}
\newcommand{\bra}[1]{\mbox{$\langle #1 |$}}
\newcommand{\ket}[1]{\mbox{$| #1 \rangle$}}

\newcommand{\auth}{AUTHORS}

\def\thb{\bar{\theta}}
\def\Thb{\bar{\Theta}}
\def\barp{\bar{p}}
\def\barq{\bar{q}}
\def\barc{\bar{c}}
\def\bard{\bar{d}}
\def\e{\epsilon}

\def \bi{\bibitem}
\def \la {\label}

\def \l {\lambda}
\def\foot{\footnote}
\def \tl  {{\tilde \l}}
\def \sql {{\sqrt \l}}
\def \adss {$AdS_5 \times S^5$\ }
\newcommand{\rf}[1]{(\ref{#1})}
\def \ov {\over}

\def\th{\theta}
\def\Th{\Theta}
\def\vth{\vartheta}
\def\btheta{{\bar\theta}}
\def\ttheta{{{\tilde\theta}}}
\def\bttheta{{{\bar\ttheta}}}
\def\vth{\vartheta}

\def\ra{\rightarrow}
\def\N{{\cal N}}
\def\F{{\cal F}}
\def\uM{\underline{M}}
\def\uN{\underline{N}}
\def\uP{\underline{P}}
\def\cc{\circ}
\def\eqv{\equiv}

\def\ni{\noindent}
\def \ha{{1\ov 2}}
\def \bw {{\rm w}}

\def\r{{\rm r}}
\def \rg{{\rm g}}

\def\Y{{\rm Y}}
\def\X{{\rm X}}
\def\tY{\tilde{\rm Y}}
\def\tX{\tilde{\rm X}}
\def\dY{\dot{\rm Y}}
\def\dX{\dot{\rm X}}

\def \J {\mathcal{J}}
\def \del {\partial}

\def\dF{\dot{F}}
\def\dG{\dot{G}}
\def\df{\dot{f}}
\def \E {{\cal E}}

\def \S {{\cal S}}
\def \J {{\cal J}}

\def\ms{\mathcal{S}}
\def\mj{\mathcal{J}}
\def\soj{\fr{\ms}{\mj}}
\def \R {{\bf R}}
\def \om {\omega}
\def \tH {\widetilde H}
\def \bE {\bar E}
\def \x {{\cal X}}

 \def \bb {\bar \beta}
\def \W {{\cal E}}

\def \bi{\bibitem}
\def \la {\label}

\def \l {\lambda}
\def\foot{\footnote}
\def \tl  {{\tilde \l}}
\def \sql {{\sqrt \l}}
\def \sqtl {{\sqrt {\tilde \l}}}

\def \HH {{\rm E}}

\def \adss {$AdS_5 \times S^5$\ }

\def \D {\Delta}
\def \thet {\theta}
 \def \t {\tau}
 \def \p {\phi}
 \def \r {\rho}
 \def \rN {{\rm N}}
 \def\tw{{\tilde w}}
 \def\hJ{{J}}
 \def\hw{{w}}
 \def\hl{{\lambda}}
 \def\hth{{\theta}}
 \def\NN{{\cal N}}
 \def \bv {{ \bar w}}
\def \vn {{\vec n}}

\def \ov {\over}

\def \varpi {{\rm w}}
\def \OO {{\cal O}}

\def\IR{\mathbb{R}}
\def\IC{\mathbb{C}}
\def\IZ{\mathbb{Z}}
\def\IP{\mathbb{P}}
\def\id{\protect{{1 \kern-.28em {\rm l}}}}

\def \rt {{\rm t}}

\def\r{{\rm r}}
\def\Y{{\rm Y}}
\def\X{{\rm X}}
\def\tY{\tilde{\rm Y}}
\def\tX{\tilde{\rm X}}
\def\dY{\dot{\rm Y}}
\def\dX{\dot{\rm X}}

 \def \bJ {{\rm J}}

\def \adss {$AdS_5 \times S^5$\ }

\begin{document}
\overfullrule=0pt
\parskip=2pt
\parindent=12pt
\headheight=0in \headsep=0in \topmargin=0in \oddsidemargin=0in

\vspace{ -3cm} \thispagestyle{empty} \vspace{-1cm}

\begin{center}

{\Large\bf Slow-string limit and ``antiferromagnetic'' state\\
\vspace{0.3cm}
 in  AdS/CFT }

 \vspace{.5cm} { R. Roiban$^{a,}$\footnote{radu@phys.psu.edu}, 
  A. Tirziu$^{b,}$\footnote{tirziu@mps.ohio-state.edu}
 and A.A.
 Tseytlin$^{c,b,}$\footnote{Also at
 Lebedev  Institute, Moscow.
  %tseytlin@mps.ohio-state.edu
 }}\\
 \vskip 0.3cm

{\em $^{a}$Department of Physics, Pennsylvania  State University,\\
University Park, PA 16802 , USA\\
$^{b}$Department of Physics, The Ohio State University,\\
Columbus, OH 43210, USA\\
\vskip 0.08cm $^{c}$  Blackett Laboratory, Imperial College,
London SW7 2AZ, U.K. }

\end{center}

 \begin{abstract}
 %%%%%%%%%%%%%%%%%%%%%%%%%%%%%%%%%
We discuss a  slow-moving limit of a rigid circular 
 equal-spin string  solution  on $R \times S^3$. We suggest that the  
  solution  with 
 the  winding number  equal to the total spin 
 approximates the  quantum string 
 state  dual  to the maximal-dimension ``antiferromagnetic'' state 
 of the  $SU(2)$ spin chain on the gauge theory side. 
 An expansion of the string action near this solution leads to 
 a weakly coupled system of a sine-Gordon model and a free field. 
 We show that  a similar effective Hamiltonian 
 appears in a certain continuum limit 
 from the half-filled Hubbard model 
 that was recently suggested to describe the  all-order dilatation
operator  
 of the dual gauge theory in the $SU(2)$ sector. 
 We also  discuss some other  slow-string solutions 
 with one spin component in $AdS_5$ and one in $S^5$.

\end{abstract}
\newpage

\renewcommand{\theequation}{1.\arabic{equation}}
 \setcounter{equation}{0}

%%%%%%%%%%%%%%%%%%%%%%%%%%%%%%%%%%%%%%%%%%%%%%%%%%%%%%%%%%%%%%%%
\section{Introduction}
%%%%%%%%%%%%%%%%%%%%%%%%%%%%%%%%%%%%%%%%%%%%%%%%%%%%%%%%%%%%%%%%%%%%5

%\ci{gkp}

One implication of the
 AdS/CFT duality  is that each free
  string state in \adss
should correspond to a certain single-trace
 gauge invariant operator in the  large $N$  maximally supersymmetric 
    SYM gauge
theory; the quantum string energy $E$ should be  equal to the
  quantum dimension $\D$ of the  operator.

  For example, if we look at the  closed $SU(2)$ sector  of operators
 like  Tr$(\P_1^{J_1} \P_2^{J_2})$  built out of two complex combinations
 of $\N=4$  SYM scalars and diagonalize the corresponding dilatation
 operator (for a review see \ci{beisrev})  then each of its eigenstates
 with given R-charges $(J_1,J_2)$
 should be dual to a particular string state with the same $SO(6)$
 spins  $(J_1,J_2)$, and one should have also 
 $\D(J_1,J_2,m;\l) = E(J_1,J_2,m;\sqrt{\l})$   \ci{bmn,gkp,ft2}.
 $\l$  in $\D$
 is the `t Hooft coupling and $\sqrt{\l} $ in $E$ is 
 the  string tension; $m$ in $\D$ labels 
various eigenstates with fixed $J_1$ and $J_2$  while $m$ in $E$ 
stands for other ``hidden''
 quantum numbers like winding number or number of folds of  the string
 configuration.

 The leading one-loop term in the dilatation operator in this  sector
 can be identified with  the Hamiltonian of the  XXX$_{1/2}$
 ferromagnetic spin chain of length $J=J_1+J_2$ \ci{mz},  while  higher-loop  corrections
 add long-range  and multi-spin interaction terms \ci{bks,beis,beisrev}.
 For small $\l$  higher  loop corrections are not expected to
 qualitatively change the structure of the spectrum of the   XXX$_{1/2}$
 model (modulo possible lifting of degeneracies), i.e. they should
  deform  the eigenvalues   order by order in $\l$.
  One may then conjecture that  the  same should be true also in   the
  large $\l$ limit, i.e. the exact spectrum should have 
  the same qualitative
  structure as the Heisenberg model spectrum.  This conjecture
  seems to be  supported,  for large  length $J$,  by the close
  relation   between the standard one-loop   Bethe ansatz
   and the exact asymptotic Bethe ansatz \ci{bds}  (see also \ci{rss}).

The AdS/CFT duality then implies that the Heisenberg model spectrum
 and  the  corresponding part of
 the quantum string spectrum should have the same
  qualitative structure.\foot{The 
  identification  of the
 ``$SU(2)$'' part of  the quantum string spectrum is, in general,
 non-trivial. 
  Here we shall assume   that at least  in the leading
  large $\l$ limit the corresponding  states  can be 
  represented by strings
  moving  in the $S^3$ part of $S^5$  with two non-zero  angular momenta
  $(J_1,J_2)$. Our main interest will be the  $J_1=J_2$ state 
  for  which  the complications discussed in  \ci{minah} seem to be
  absent.}  
 The spectrum  of the one-loop  ferromagnetic Heisenberg model
 ($ \D= J + E_{(1)}, \ \ E_{(1)} = \l F_1(J_1,J_2) $)
 starts with  the  ground state represented by the  
 BPS operator Tr $\P^{J}_1$
 and
 dual  to the point-like  string moving along geodesic of   $S^5$.
 Small  fluctuations
 near the ground state, i.e.  magnons,   with
  $E_{(1)} \sim  { \l \ov J^2} $,  
 are represented, at large $J$,   by the  BMN \ci{bmn}
 operators (with $ J_1 \gg J_2 $);
  extrapolated to large coupling they are dual
   to small  strings rotating within $S^3$ part of $S^5$
   with their  center of mass moving along the   big circle.
   
 States with higher energy  may be described  as  bound states of magnons
 or
  ``macroscopic strings'' of Bethe  roots  \ci{haldchas,bmsz}. In
  the thermodynamic limit with
 $J_1 \sim J_2 \gg 1$  one finds,  using the Bethe ansatz, 
  that for them
  $E_{(1)} \sim  { \l \ov J} f({J_1\ov J_2})  $ \ci{bmsz}.
   They may be  represented by ``locally-BPS''
    Tr$(\P_1^{J_1} \P_2^{J_2})$ -type operators with
   slowly changing  order of   $\P_1$ and $\P_2$ clusters.
 The strong-coupling extrapolation  of these states
 are ``fast'' 
 ($J_1 \sim J_2$, \ $ J \sim \sqrt \lambda  \gg 1$)
  semiclassical strings
     whose world surface is approximately null
 \ci{ft2,ft4,mmt,mikh,kru,kt}.

 Going higher in the energy, the  spectrum is expected to contain,  
for $J \gg 1$, some   ``intermediate'' states with 
 $E_{(1)} \sim \l  + O(1/J) $
 and,  finally, 
  the highest energy state with 
 $E_{(1)} \sim \l J  + O(1) .$
  In the latter  case the  energy density
 will  be approximately constant at large $J$  instead of vanishing
 as for the  magnons or ``macroscopic strings''.
 
   Indeed, the spectrum of the ferromagnetic Heisenberg chain
   is isomorphic  to the spectrum of the antiferromagnetic chain:
   the two spectra   are formally
   related by changing the sign of the overall coefficient 
   $\l$ or the sign of
   the energy.\foot{Changing formally the sign
   of $\l$ in the full all-loop dilatation operator
   will not, of course,   lead to an
   antiferromagnetic chain with  isomorphic spectrum;
   for example, the BDS  ansatz \ci{bds}
     has non-trivial dependence on $\l$
   through the magnon dispersion relation $e(p) = \sqrt{ 1 + {\l \ov
   \pi^2} \sin^2 { p\ov 2} }-1 $ (cf. also \ci{rss}).
   Still, as already mentioned above, the
    exact spin chain Hamiltonian
   is expected to have the  spectrum
   (including its  higher-energy  near-antiferromagnetic-state part) 
    which is a smooth
   $\l$-deformation of the Heisenberg  model spectrum.}
    This implies that the highest-energy state  in the
   Heisenberg  ferromagnet spectrum is the same as the Neel-type
   antiferromagnetic (AF)  ground state  of the
   Heisenberg  antiferromagnet, i.e. it should
    have  $J_1 =J_2=J/2$ and  for $J \gg 1$ its energy should be 
    \ $E_{(1)} =  c_1  \l   J $,\ \  $c_1 = { \ln 2 \ov 4 \pi^2} $\
    \ci{frad,fadd}.
   The fluctuations near  the  AF state  
   will  lower the  energy of the ferromagnetic chain,
   eventually filling up  the
   part of the spectrum from  the near-AF states
     with $E_{(1)} \sim \l J $
   to the ``intermediate'' states with $E_{(1)} \sim \l $.

 Beyond the  1-loop order one expects to find that the
 energy of the AF  state  should be given  by  (assuming
 $J \gg 1$ for any fixed $\l\ll 1 $)
 \be \la{gen}
 E= f(\l)\ J \ , \  \ \ \ \ \ \ \
 f(\l \ll 1) = 1 + c_1 \l + c_2 \l^2 + ... \ . \ee
 The exact expression  for the ``energy density'' $f(\l)$
 was recently found by starting with the conjectured asymptotic
 BDS \ci{bds}  Bethe ansatz \ci{rss,zar}
 \be \la{bes}
   f(\l)
=1+\frac{\sqrt{\lambda }}{\pi }\int_{0}^{\infty }
 \frac{dk}{k}\,\,\frac{\bJ_0\left(\frac{\sqrt{\lambda }}{2\pi }k\right)
 \bJ_1\left(\frac{\sqrt{\lambda }}{2\pi }k\right)}{\,{\rm
 e}\,^{k}+1}\ . \ee
Here $\bJ_n$ are the Bessel functions.
 The formal extrapolation of this expression to large $\l$
gives \ci{zar}:
\be \la{ext}
 f(\l \gg 1) =\frac{\sqrt{\lambda }}{\pi ^2} + { 3 \ov 4}  + 
 ...
 %{\rm
 %exponential corrections}
 \ .
\ee
The BDS ansatz   is  not expected to
 correctly  represent the quantum string spectrum at the  quantitative
 level,  
 %(so the coefficient of $\sql$  may not be the same as 
 %found from the
but   it  was previously found to lead to the  same qualitative results
 for its  low-energy  part.\foot{In particular,
 the low-energy Landau-Lifshitz-type effective actions
 corresponding to the  BDS ansatz  and the quantum string theory
 appear to have the same structure \ci{krt,mtt2}.}
 The same is likely to be true also for the upper part of the spectrum
 (in the large $J$ limit). 
 Indeed, the arguments in \ci{zar}
 suggest  that 
 one should find the same $f(\l \gg 1) 
\sim \sql $ scaling by starting with the string Bethe ansatz of \ci{afs} 
(though the proportionality coefficient is likely to be different 
from $1/\pi^2$).

 It is then natural to conjecture 
  that  the energy of the string state
 dual  to the AF state  should scale
 in the same way  at  large $\l$
 (i.e. in the classical string limit)
   and  large $J$, i.e. we should find  
 \be \la{str}
   E (\l \gg 1) \sim \sql\  J  \ . \ee
   This is  {\it not}  the  familiar 
    scaling for  semiclassical strings
   of the type discussed in \ci{gkp,ft1,ft2} (for which
   both the energy $E$ and the spin $J$ of the classical
   string  are proportional to  the  string tension, i.e. 
   scale as $\sql$ at large $\l$) so one may then question if the AF state
   can be represented by a semiclassical string.

Since the string should carry the large  angular momentum
  $J$  one may hope that the corresponding quantum string state
  may still be approximated -- in the large $\l$ limit --
  by a classical string configuration.
  
  \bigskip\bigskip
 Our aim below will be  to try  to identify
 semiclassical string states that should
 be dual to the upper part
 of the  gauge-theory spin  chain spectrum in the
 limit of large $J$ and large $\l$. We shall find 
 that there are  indeed  classical string  solutions 
 whose energy   scales as \rf{str}.\foot{Related question
 was already studied  in \ci{ptt1}  for
 2-spin string states in $AdS_5$  dual to  gauge-theory operators built
 out of  self-dual part of gauge field strength
 whose 1-loop anomalous dimensions are described by antiferromagnetic
 XXX$_1$  spin chain \ci{hfz}.}
  However, 
  the semiclassical expansion  here will have an unusual form, with
  subleading terms in the classical  energy receiving contributions from
  higher orders  in string $\a'\sim { 1 \ov \sql}$  expansion.
  Also, the  classical string solution  will be    unstable
  under small fluctuations.
  That direct semiclassical expansion may not
  apply here  is  not surprising since it is  only the
   true quantum string state
  that should be dual to the AF state on the gauge theory state.
  While the AdS/CFT duality implies that
   the   quantum string theory spectrum,
   being equivalent to the gauge spin chain spectrum,
  should be bounded from the  above  in the
 compact $SU(2)$ case   \ci{zar,rss},  adding
 small fluctuations to a semiclassical string
 one can always increase the  energy.

 Our main observation is that while the lower part of the $SU(2)$
 spin chain spectrum  is   dual to  {\it fast-moving}  strings
 (which are ``locally null-geodesic''  or ``locally BPS''),
      the  upper part   appears to be dual to
 {\it slow-moving long} strings which are
 as far as possible  from the BPS limit.
 While for the  fast strings   the time ($\tau$)
 evolution of the string
 configuration   dominates  over the spatial ($\s$) evolution, 
 with each bit of the string having a near-null-geodesic trajectory,
 for the slow long string one has just the opposite:
  each of its bits
 moves very slowly.  The slow  motion is  not in contradiction with the
 assumption that $J \gg 1$: the  effective string rotation frequency
 $\J  = {J \ov \sql}  $  is very small
 in the classical string limit $\sql \gg 1$
 if we assume  that $ \sql \gg J$.
 This should be contrasted  with the fast string case  where
 $\J $ was fixed in the limit  of $\sql \gg 1$,
  i.e. $J$ and $\sql$ were of the same order \ci{ft2}.
  \foot{There the effective coupling $\tl = {1\ov \J^2} =
  { \l \ov J^2} $ was fixed 
  while $\l$ was taken large. One could then expand  in
  powers of $\tl$
   at each order of semiclassical expansion in $1\ov \sql$.}

 We  propose   that the quantum string state  representing the
 AF state of the gauge theory  may be approximated by
 the simplest rotating  string
 solution in $S^5$  \ci{ft2,art}:
 a circular   string  moving  in $S^3$ part of $S^5$
 with two equal angular momenta $J_1=J_2=J/2$  and 
  wound along a  big circle. The winding number $m$ should be equal to the
 total momentum $J$, i.e. the total length of the string in $\s$
 direction should be proportional to $J$.
 
 The assumption that  the winding number $m$ should be proportional to the
 angular  momentum $J$ is a natural one from
 the spin chain/Bethe ansatz   point of view:   for  the  AF state  case
 the excitation  momenta $p_i$ determined by the Bethe ansatz
  and thus the energy density $E/J$ (with $E= \sum^{J_2}_{i=1}
  [ \sqrt{ 1 + {\l \ov
   \pi^2} \sin^2 { p_i\ov 2} }-1 ]$)
  should be   constant  in the large length $J$ limit
 (similar  limit was considered  in \ci{polman}).
%%foot{
%% %REV
%% In fact, the
%% Bethe root momentum
%%  distribution for   the antiferromagnetic state
%%  considered in the context of the BDS ansatz  in
%% \ci{zar} was  such that the corresponding winding number
%% $m = { 1 \ov 2 \pi} \sum^M_{k=1} p_k$  (with $M= L/2$)
%%  was of order $L$  at   strong coupling (for large $L$).
%% }
%\foot{
% %REV
% In fact, the 
% Bethe root momentum 
%  distribution for   the antiferromagnetic state
%  considered in the context of the BDS ansatz  in 
% \ci{rss,zar} was  such that the corresponding winding number
% $m = { 1 \ov 2 \pi} \sum^M_{k=1} p_k$  (with $M= L/2$) 
%  was of order $L$ both at weak  and  strong coupling.
%  At strong coupling \ci{zar} $p_k = { 4 \pi k \ov L}, 
%  \  \ k= 1, ..., L/2$ so that 
%  $m= { 1 \ov 4} (L +2) \to { 1 \ov 4} L$ at large $L$.
%  }
 The classical energy  of the circular string \ci{ft2}
 $E= \sqrt{ J^2 + m^2 \l }$   with $m=J$  becomes
 $E= \sqrt{ 1 +  \l } J = (\sql  + { 1 \ov 2 \sql} +  ...) J $.
 Here  the first term in the large  $\l$ expansion
   is indeed the same as in \rf{str}.
    Only this leading term
   in the large  $\l$ expansion of the classical energy
 should be trusted since the  subleading terms
 will receive contributions from higher  quantum string corrections
 (see below).

 A qualitative reason for the existence of the ``slow'' string states
 is the  compactness of $S^5$:  in flat space the closed string needs
 to rotate or pulsate
 to balance  its  tension, while  on a  sphere it can
 be wrapped on a  big circle  and thus can be static
 (to embed such a state in the $SU(2)$ sector
 we need still  to add two angular momenta and take $J$ large).
 The apparent small-fluctuation
   instability  of the wrapped (and rotating)  classical string 
    solution \ci{ft2}  may  be interpreted as  an indication
   that the  corresponding 
   quantum state has the maximal energy for  
   given spins $J_1=J_2$.

% by the way , there is actually a slow folded string state whose energy goes as
% c sqrt lambda  -- this is a folded string at rest  with only one component
%of the angular momnetum -- discussed in GKP
%-- it should probably correspond to a stte in the middle of the spin
%chain spectrum -- transition from 1/J to J  scaling
%Could you look more carefully at this case.
%I worry though that it may not be  in the SU(2) sector as we lack J_2
 %the semiclassical expansion will be unusual

%similar identification of AF state with circular pulsating
%solution was suggested in ...

\bigskip 
The rest of the paper is organized as follows. 
In section 2 we shall review  the circular two-spin solution and consider
its fast and slow  limits. 

%
% new
%
In section 3  we shall   discuss
% 
%by analogy with the
%Landau-Lifshitz-type action 
%in the fast string limit \ci{kru}, 
the 
effective Hamiltonian for  the fluctuations 
%in the slow string limit
around the  string state corresponding to the AF state of the gauge
theory chain. 
It is   found by expanding the string Hamiltonian near the circular string solution 
and is expected to be related to a strong coupling limit of an
effective action 
describing  fluctuations near the AF state  of the gauge-theory spin chain.
In contrast to the XXX$_{1/2}$ case this Hamiltonian need not be just that
of a relativistic sigma model on $S^2$,  but may be related to 
a bosonized field theory limit of a Hubbard-type model that may 
represent the all-loop dilatation operator of gauge theory in 
the $SU(2)$ sector 
\ci{rss}.

Indeed, we will show in section 4   that  a similar effective Hamiltonian 
 appears in a  continuum limit from the half-filled Hubbard model. 
The bosonized Hamiltonian exhibits a certain discontinuous behaviour
as we move away from half-filling. This bears certain similarities
with the non-closure  of the $SU(2)$ sector at large coupling 
for states with
$J_1\ne J_2$  \ci{minah} suggesting that the Hubbard model may need
 to be modified to
take this into account.

%1. take XXX_1/2 and get that strongly coupled dndn action a la Fradkin et al
%2. try redoing their analysis   with say 2-llopp or 3-loop terms added
%to dilop as H
%what do we get ? in the case of LL  we needed quantum corrections also
%to be added but at least <H>  gave an indication of higher derivative
%terms to be included
%3.. now compare that to prediction of Hubbard...

In section 5 we shall consider a similar slow string limit of some
string  solutions with one spin in $AdS_5$ and one spin in $S^5$ that 
are related, in particular, to the $SL(2)$ sector on the gauge side. 
Quantum 1-loop correction to the energy of 
the latter  solution will be discussed in Appendix.  

%A limit of large winding number 
 %was also considered in \ci{szz} in the case of the circular  $SL(2)$ 
 %sector solution of \ci{art}. 
 %We shall discuss this solution and its possible  limits in 
 %section 4 below.

%%%%%%%%%%%%%%%%%%%%%%%%%%%%%%%%%%%%%%%%%

%%%%%%%%%%%%%%%%%%%%%%%%%%%%%%%%%%%%%%%%%%%%%

\renewcommand{\theequation}{2.\arabic{equation}}
 \setcounter{equation}{0}

\section{Circular rotating $J_1=J_2$ string  on  $S^3$}
%%%%%%%%%%%%%%%%%%%%%%%%%%%%%%%%%%%%%%%%%%%%%%%%%%%%%%%

Here   we shall start  with recalling the form of the simplest 2-spin
solution  \ci{ft2} for the string on $R_t \times S^3$
 (in the form given in \ci{art})
  and then discuss  its 
new ``slow-string''  limit. 

\subsection{Classical string energy and its limits}

Parameterizing $S^3$  by two complex coordinates $\XX_i$
with  $ |\X_1|^2   + |\X_2|^2  =1$,  the classical string equations
 in conformal gauge  may be
 written as $\del^2  \X_i + \L \X_i =0$, where 
 \ $ \L= \del^m \X^*_i \del_m X_i.  $
One finds the following solution 
 \be \la{hhh}
 t= \k \tau \ , \ \ \ \ \ \
  \quad \X_1=a  \ e^{iw\tau+im\sigma}\ , \quad\ \ \
 \X_2=a \ e^{iw\tau-im\sigma}\ , \ \ \ \
  a= {1 \ov \sqrt 2} \ ,
 \ee
 where $m$ is an integer winding number (we
 shall choose it to be  positive). 
 Note that a similar solution exists in flat space
 where $w=m$ \ ($\L=0$) and $a$ is arbitrary.
 The conformal gauge constraint gives
 $ \kappa^2=w^2+m^2$.
 The  corresponding $SO(4)$ spins and the energy are ($T= { \sql \ov 2 \pi}$
 is the string tension)
 \be
 J_1 = J_2 = J/2 \ , \ \ \ \  \ \  J = \sql w \ , \ \ \ \ \ee
 \be \la{ene}
 E= \sql \k  = \sql \sqrt{ w^2+m^2}
 =\sqrt{J^2+\lambda m^2} \ .
 \ee
The quadratic
fluctuation spectrum near this solution
was found in  \cite{ft2,art}.
There are 4  massive $AdS_{5}$
fluctuations  with
mass $\kappa$, i.e. with the characteristic frequency
$\omega_n =\sqrt{n^2+\kappa^2}=\sqrt{n^2+w^2+m^2}$.
 In addition,  there is a free massive field from $S^{5}$ with
mass $w^2-m^2$, i.e. with
$\omega_n =\sqrt{n^2+w^2-m^2}$ which  is real if
\begin{equation}
n^2+w^2-m^2\geq 0 \ . \label{ffreq}
\end{equation}
The remaining three  $S^{5}$ bosonic fluctuations are coupled
and the corresponding  frequencies are given by \ci{art}
\begin{equation}
\omega^2_n=n^2+2w^2\pm 2\sqrt{w^4+n^2 w^2+m^2 n^2}\ .
\end{equation}
Their reality condition is 
\begin{equation}
n\geq 2m \ ,   \label{cfreq}
\end{equation}
which, if satisfied, implies also \rf{ffreq}.
As a result, there  is always a finite number of unstable modes  with 
$0 < n < 2m$, i.e. the solution is always unstable.

Returning to the  classical
 energy, we see that it  is a function of three independent
parameters: $\l, J, m$.
Taking
different limits of these parameters one  finds special cases 
of this  solution that
have  different physical interpretation.

Let us  first consider the  cases where the  
 standard  semiclassical expansion applies, which assumes
 that the parameters of the solution $w,m$
 are fixed while  $\l$ is taken to be large to  suppress
 quantum string  (inverse tension) corrections.
 Then $J= \sql w $ is also large, and $J \gg  m$.
 
 There are several possible choices of 
 the rotation velocity $w$ and the winding number $m$:

 (i)\  $m=0, \ w\not=0$: \ \ \ \ this is the point-like (BPS) case 
  with $E=J$.\foot{The corresponding massless geodesic 
  runs along  big circle in the ``diagonal'' 2-plane in 
  $(\X_1,\X_2)$ space.}

 (ii)\  $m \ll w$:  \ \ \ \ this is
 the ``fast string''  case of  \ci{ft2}    when $E$
 has  a regular expansion in the small semiclassical parameter
  $ { m^2   \ov w^2}= m^2 \tl$, \ $\tl \equiv  {  \l  \ov J^2} $:
 \be E= J \sqrt{ 1 + m^2 \tl}= J  (1 +   \ha m^2 \tl + ... ) \ee
 Here the time evolution is dominating over the   spatial evolution:
 the effective string tension  $ {  m \sql  \ov J} $
 is small.
 %This solution has finite number of unstable modes.

 (iii) \  $m = w$:\ \ \ \
 such   solution is formally the same  as in flat space
 (the  Lagrange multiplier $\L$ vanishes), but
  the classical energy  here is still linear in $J$:
  \be
E=\sqrt{2}\  J \   \ee
%The  modes with  $ 0 < n < 2m$ are still tachyonic so the solution
%is unstable.

(iv)  \  $m \gg  w$:\ \ \ \ here $\k$ or $E$ can be
expanded in $ {w \ov m } \ll 1$  getting
\begin{equation}
E=\sqrt{\lambda}m \sqrt{ 1 + {J^2 \ov m^2 \l} }
= \sqrt{\lambda}\ m + \frac{J^2}{2m \sqrt{\lambda}}+...\ .
\end{equation}
$m$ may  be of order 1  or much bigger than 1
but is still  much smaller than $J= \sql  w$
since $\l$ is assumed to be  taken to be  large first.\foot{
%REV
The scaling 
of the energy of long wound strings with winding $E= \sqrt{\lambda} m  +...$
was observed  in the uniform 
gauge Hamiltonian formalism  in \ci{af1}; 
however, in contrast to  \ci{ft2} and the present discussion, 
 the winding  there  was assumed  to be in the same
 direction as the momentum $J$. 
 Similar behavior is found  also in the $su(1|1)$ sector which 
was  analyzed in detail  (for any $J$ and $m$)  in \ci{af2}.
 }

The cases (iii) and (iv) are different from the
 fast-moving string case (ii)
where string  world-surface is nearly null.
In the ``slow string''
case of (iv) the $\s$ dependence dominates over
$\tau$ dependence, and a reflection of that  is the explicit
dependence
of the classical energy on the string tension
$\sql$ (in  the fast string case the classical energy
depends only on the square of string tension, i.e. is analytic in $\l$
 \ci{ft2}).
Such slow strings should correspond to an 
intermediate part of the spin chain spectrum
where the  energy scales as ($J \gg 1$)
\be\la{inti}
E = f(\l)  + O( { 1 \ov J}) \ , \ee 
\be  
f(\l\ll 1 ) = a_1 \l + a_2 \l^2 + ... \ , \ \ \ \ \
f(\l\gg 1 ) = b_1 \sql + b_2 + ...\ .
\ee
 To sum up, 
 fixing  the spins 
 $J_1=J_2=J/2$ we
   may  label the  string or the corresponding
   spin chain states by  growing 
 values of $m$;  then the energy increases
 
 from  $E(m=0)= J$ 
 
 to $E( m \ll { J \ov \sql})
 =J + { m^2 \l \ov J^2}+ ...$
 
 to  $E(m= { J \ov \sql})= \sqrt 2 J $
 
 to  finally to  $E(m  \gg {J \ov \sql})
 = \sql m + ... $.

 One may wonder what will happen if we increase $m$
 or the string length further.
 The spin chain correspondence  suggests that the highest possible value
 of $m$ should be $J$ (which takes integer values in the quantum theory).
 If we assume that\foot{One may of course set 
 $m = k J$   where  $k$ is an integer, but we expect that $k>1$ cases will be
 equivalent to $k=1$ in the exact quantum theory.}
 \be
 {\rm (v)}: \ \ \ \ \ \ \ \ \ \ \  \ \ \ \  \ \ \ \ \ \ \ \ \ \ \  \ \ \ \ m = J \ \   \gg 1   \ , \ee
  but still $  m \ll \sql$ then
 $w= { J\ov \sql} ={ m\ov \sql} \ll 1  $.
 Hence in this case  the string motion
 is {\it ``very slow''}  for  large tension:
 the string  wrapped many times on  big circle is nearly static
 in the classical $\sql \gg 1$ limit.
The  energy  \rf{ene} for $m=J$ is then 
\begin{equation}
E=J\sqrt{1+\lambda} \ , \ \ \  {\rm   i.e.} \ \ \ \ \ \
E(\l \gg 1) =   \sqrt{\lambda}  J + ...  \ .
\label{cla}
\end{equation}
Our  main conjecture is that this 
special  case of the circular string solution  should  be dual
to the  highest-energy antiferromagnetic state
of the corresponding gauge-theory spin chain.\foot{Another known 
 solution in the $SU(2)$ sector
 with $J_1=J_2$ is the folded string one  \ci{ft4}. One may   wonder
 if this solution also admits a  limit when  the number 
 of folds $m$ becomes large together with $J$. 
 The answer is no: here  one cannot take  the string rotation velocity to 
 be small without having the string  shrinking to a point.}

Like in the previous cases (i)--(iv)  the solution 
in the case (v)  is still  unstable,   with the number of
tachyonic  modes with $n < 2m=2J  $  growing with $J$.
This instability  may, however, 
 be  an artifact of the   naive
semiclassical expansion  near the highest-energy state:
 our conjecture implies  that
there is  a well-defined maximal-energy state in the
discrete quantum string spectrum which in the large $\l$
limit  may be  approximated by the above classical solution
with $m=J$.

%%%%%%%%%%%%%%%%%%%%%%%

The standard semiclassical expansion does not indeed directly
apply in the last   case (v):
 the classical energy depends on $\sql$ and
contains subleading
terms that appear also from higher orders in inverse string
tension expansion (see next subsection).
Still, the leading $ \sql J$
term in the  classical string energy \rf{cla} 
 does not receive
corrections from higher world-sheet loops,   and this leading scaling behavior 
thus 
provides
a  qualitative support to  our conjecture  that this
solution (v)  is dual to the highest-energy  AF state  of the
gauge-theory   spin chain.

The fact that 
the proportionality coefficient 
$1/\pi^2 $ in  \rf{ext} 
as obtained in \cite{zar} by extrapolating to strong coupling 
the AF energy of the BDS spin chain 
does not match the  one in \rf{cla}  may not be considered as 
a contradiction. Indeed,  the orders of limits taken are opposite:
on the string side we first take 
 $\l$ large and then $J$ large, while on the gauge side 
we  first assume that $J$ is large and then extrapolate the 
perturbative in  $\l$ result to strong coupling.

%%%%%%%%%%%%%%%%%%%%%%%%%%%%%%%%%%%%%

\subsection{1-loop correction to string  energy}
%%%%%%%%%%%%%%%

Let us now consider the slow-string limit  of the 1-loop correction to the energy 
of the above circular solution which was computed in \ci{ft3,fpt} (see also \ci{btz}).
Its  expansion that was discussed before was the fast string limit 
when $w\gg m$ (for a discussion of  subtleties in this expansion see \ci{bt,sz}).
Here we shall consider the opposite limit of  $w\ll m$.
We shall formally  ignore the instability of the solution, 
concentrating on the real part of the 1-loop correction $E_1(m,w)$.
The expansion in powers of ${w\ov m} ={ J \ov \sql m } $
produces powers series in $1\ov \sql$ for fixed $J$ and $m$ (in particular, for $J=m$). 
As we shall see, for large $m$ the one-loop correction   $E_1$ will 
scale linearly with $m\gg 1 $ or, for  $m=J$, with $J$, 
in agreement with the general expectation 
that 
\be E= f(\l) J\ ,\ \ \ \ \ \ \ 
 \ f(\l\gg 1) = a_1 \sql + a_2 + { a_3\ov \sql} + ...\ . \ee 
 The  expression for $E_1$ is given by 
 the sum of the zero-mode and non-zero-mode contributions 
\begin{equation}
E_{1}=E_{\rm zero}+E_{\rm  non-zero}\  , \quad \quad
E_{\rm  non-zero}=\sum_{n=1}^{\infty}S_{n} \ ,
\end{equation}
\begin{equation} \la{kkk}
E_{\rm zero}=2+\sqrt{1-\frac{2m^2}{w^2+m^2}}-3\sqrt{1-\frac{m^2}{w^2+m^2}}\ , 
\end{equation}
\begin{eqnarray}
S_{n}&=&2\sqrt{1+\frac{(n+\sqrt{n^2-4m^2})^2}{4(w^2+m^2)}}+2\sqrt{1+\frac{n^2-2m^2}{w^2+m^2}}+
4\sqrt{1+\frac{n^2}{w^2+m^2}}\nonumber\\
&-&8\sqrt{1+\frac{n^2-m^2}{w^2+m^2}}\ . 
\end{eqnarray}
In contrast to the large $w$ expansion  relevant for the fast string case, the small $w$ expansion 
of the 1-loop correction  is regular:  higher orders coefficients are given by convergent sums. 

The leading term  in the expansion in $w/m$ is found by setting $w=0$ in the above expression
(omitting the imaginary part):
\be 
E^{(0)}_1=2 + { 1 \ov m} \sum_{n=1}^{\infty} \bigg[
\sqrt{4 m^2 +  (n+\sqrt{n^2-4m^2})^2}+2\sqrt{n^2-m^2}+
4\sqrt{n^2 + m^2} -8 n \bigg] \ . \ee 
The expansion of this at large $m$ is subtle but numerical 
evaluation shows that the real part of $E^{(0)}_1$   scales linearly with $m$ at large 
$m$, supporting the suggested identification of the corresponding string solution with a 
 particular $J_1=J_2$ ``intermediate'' state of the spin chain. 

Setting $m=J=\sqrt{\lambda}\ w $ in \rf{kkk}  we get 
\begin{equation}
E_{1}=E_{\rm zero}(\lambda)+E_{\rm  non-zero}(J,\lambda) \ ,
\end{equation}
\begin{equation}
E_{\rm zero}=2+\sqrt{\frac{1-\lambda}{1+\lambda}}-3\frac{1}{\sqrt{1+\lambda}},
\quad \quad E_{\rm  non-zero}=\sum_{n=1}^{\infty}S_{n}(J,\lambda)
\end{equation}
To analyze the dependence of $E_{1}$ on $J$, we expand $S_{n}$ at
large $\lambda$ for fixed $J$:
\begin{equation}\la{ji}
S_{n}= { 1 \ov J} ( A_{0}+\frac{A_{1}}{\lambda}+\frac{A_{2}}{\lambda^2}+...)\ , 
\end{equation}
\begin{equation}
A_{0}=-8n+2\sqrt{n^2-J^2}+4\sqrt{n^2+J^2}+\sqrt{2n(n+\sqrt{n^2-4J^2})}
\end{equation}
\begin{equation}
A_{1}=4n-\frac{4J^2}{n}-\frac{2n^2}{\sqrt{J^2+n^2}}+\frac{2J^2-n^2}{\sqrt{n^2-J^2}}+
\frac{2J^2-n(n+\sqrt{n^2-4J^2)}}{\sqrt{2n(n+\sqrt{n^2-4J^2})}}
\end{equation}
\begin{eqnarray}
A_{2}&=&\frac{J^4}{n^3}+\frac{2J^2}{n}-3n-\frac{n^4}{2(n^2+J^2)^{3/2}}
+\frac{(4J^4-8J^2n^2+3n^4)\sqrt{n^2-J^2}}{4(J^2-n^2)^2}\nonumber\\
&+&\frac{2n^2}{\sqrt{n^2+J^2}}+\frac{3n^3
(n+\sqrt{n^2-4J^2})-2J^4-2J^2n(4n+\sqrt{n^2-4J^2})}{2\sqrt{2}[n(n+\sqrt{n^2-4J^2})]^{3/2}}
\end{eqnarray}
%Note that here a potential issue with the large $\lambda$
%expansion does not appear since the series $\sum A_{n}$ are found
%to be convergent.\footnote{This is expected here since, as we
%already pointed out, taking large $\lambda$ limit with $J$ fixed
%is identical to taking small $\mathcal{J}$ limit with $\lambda$
%fixed. But taking the latter limit in the sum of the 1-loop energy
%$\sum S_{n}(\mathcal{J},\lambda)$ presents no problem.}
To study the $J$-dependence of the series we computed the sums 
 $\sum_{n=2J}^{N}A_{0}$,
$\sum_{n=2J}^{N}A_{1}$, $\sum_{n=2J}^{N}A_{2}$ 
numerically 
for  $N=10^5$ and  $10^2<J<10^4$. We found  that they scale as $J^2$, 
so that $S_n$ in \rf{ji} grows linearly with $J$. 

Numerically evaluating the coefficients
  and combining 
$E_1$ with the classical expression \rf{cla} we get 
for the large $\l$ expansion of $E= E_0 + \hbar E_1 + ...$
 (ignoring $O(J^0)$ terms in $E_1$, i.e.,  in particular, 
 terms coming from $E_{\rm zero}$): 
\begin{equation}
E=J\bigg[\sqrt{\lambda}\bigg(1-
0.34\frac{\hbar}{\sqrt{\lambda}}\bigg)+
\frac{1}{\sqrt{\lambda}}\bigg(\frac{1}{2}+
0.215\frac{\hbar}{
%4.6
\sqrt{\lambda}}\bigg)-\frac{1}{\lambda^{3/2}}\bigg(\frac{1}{8}+
0.16\frac{\hbar}{%6.3
\sqrt{\lambda}}\bigg)+...  \bigg] \ . 
\label{qenergy}
\end{equation}
Here we formally  introduced the 
parameter $\hbar$  to distinguish between the 
classical and the 1-loop corrections. 
It is interesting to observe that  while the classical 
part of $E/J$ contains $1\ov \sql$ terms in odd powers,  the  1-loop
corrections  
produce the even  powers of $1\ov \sql$.
The subleading coefficients will be further corrected by higher loop  string
contributions. 
This illustrates 
the point that was already mentioned above: in contrast to the usual
semiclassical 
expansion  in the $m=J$ case 
the string sigma model loop expansion is not equivalent to large $\l$
expansion.

It is interesting also to note that the leading $-0.34{\hbar}$
correction  
to the classical $\sql$ term in $E/J$ is negative, which seems
consistent with the idea of interpolation from strong to weak coupling
(cf. \rf{gen}).  

Returning to the issue of instability of the solution,  we expect that 
it is related to the fact that one tries to expand near a maximum of a
potential  
like $\sin^2 \theta$. The exact quantization should produce a discrete
set of levels  
in this potential 
with the maximal energy state being ``approximated'' by the above
classical solution.\foot{  
It is possible also that the relevant quantum string state may  be
better represented by a  
pulsating string \ci{gkp,minahan,emz,kt} 
(with pulsations outside of $S^3$). Given that a precise meaning 
of the closure of the $SU(2)$ sector is unclear on the string side
(cf. \ci{minah}),  and, 
in particular, that the quantum string fluctuations ``feel'' all
directions of \adss, 
it is possible that there exists a pulsating string solution which for 
large $J$ (bigger than its oscillation number)  has  essentially the same 
energy as the unstable rigid rotating string we discussed here.}

%%%%%%%%%%%%%%%%%%%%%%%%%%%%%%%%%%%%%%%

\renewcommand{\theequation}{3.\arabic{equation}}
 \setcounter{equation}{0}

\section{Effective action for slow-moving strings on $R \times S^3$}
%%%%%%%%%%

In the case of the  lower part of the spectrum of 
the ferromagnetic 
spin chain dual to fast strings   it was possible  to establish a correspondence
 between 
a non-relativistic  Landau-Lifshitz (LL)    effective action for  long-wave
 length excitations 
of the spin chain and the  fast string limit of the classical string action 
\ci{kru,krt,mih,kt}. One may wonder if a similar kind of 
 effective action correspondence 
exists also  near the upper antiferromagnetic end of the spin chain
spectrum  related to a slow-string limit 
of the string action. 

As is well known (see, e.g., \ci{frad,sach}),  the effective action
describing  
near AF-state excitations of the XXX$_{1/2}$  spin chain is a relativistic 
sigma model on $S^2$ (with a topological term ensuring its conformal
invariance  
and no-gap spectrum). The exact  spin chain  representing 
 gauge theory
anomalous  
dimensions is  certainly  different  from the 
XXX$_{1/2}$  chain  and the strong-coupling limit of the corresponding 
effective action need not be simply an  $S^2$  relativistic sigma model
as in the XXX$_{1/2}$  case.
The exact 
 spin chain 
was suggested  to be related to a version  of the 
 Hubbard model \ci{rss}.
It is not completely 
clear at the moment which is the correct Hubbard-type
model   
which should be related  to 
string theory and which should be the corresponding 
near-AF state effective action for it, but one may
 assume that it should be 
qualitatively similar to that of the Hubbard model.
The  continuous effective action for the fluctuations 
near the ground AF state  of 
 the half-filled Hubbard model 
is a combination of the massless spinon sigma model  and a sine-Gordon
action for 
massive  charge excitations \ci{korep,frad,tsvel}. 

Here 
 we shall first  attempt to see what kind of  effective 
 Hamiltonian  for near-AF state  fluctuations 
may appear in the dual limit on the  string side.
Then in section 4 we will find  that the form of this Hamiltonian
 and thus its spectrum 
is qualitatively 
similar to that of the Hamiltonian 
appearing in a scaling limit of the 
 Hubbard model of \ci{rss}.

%to relate the two effective actions in a precise way 
%as was done in the ferromagnetic/fast-string case. 

In general, one would  need 
 to start with the full 
quantum string theory and integrate out all modes  but the  ones  
relevant for the description of the near-AF states 
of the $SU(2)$ sector. 
Here we shall suppress  
world sheet 
quantum 
%string 
corrections
by  assuming that $\l \gg 1$, i.e. we shall 
consider only the classical string action. 
We shall follow a naive approach that essentially copies 
the derivation of the LL action in \ci{kru,krt,kt} 
but now focuses on  modes close to the wrapped slow-moving  string that 
we conjectured above  to be  the counterpart  of the AF state. 
%AATnew%
More precisely, the 
 analogy here will be with the action of magnons as 
small fluctuations near the ferromagnetic state, or with the 
corresponding ``plane wave'' action  on the string side.

%%%%%%%%%%%%%%%%%%%%%%%%%%%%%%%%%%%%%%%%%%%%%%%%%%%%%%%%%%%%
%\section{  Comments}
%
%\foot{
%The effective action  for low wave-length excitations  near the ground state
%of 1-d  antiferromagnet with Neel order  (see, e.g., 
%\ci{sach,frad})
%gives an apparently strongly coupled
%sigma model on $S^2$:  $L= {1\ov 2 k} (\del \vec n)^2 $, \ $k = 2/s=4$.
%Topological WZ term can be ignored on $R \times S^1$ ?
%In any case, we should not treat this model as quantum.
%critical model for s=1/2.
%But in usual ferromagnet  exact solution
%is \ci{Fateev-Zamolodchikov}: 
%spectrum
% is massless with spin 1/2 -- spinons (see \ci{tsvel}).

Given a classical string moving on $R_t \times S^3$  we are to gauge fix two 
coordinates (time and a spatial one) to get an action for two physical transverse degrees of freedom. 
In \ci{krt} this was done 
  by fixing the momentum density corresponding to the 
sum$\a$  of the two polar angles ($\phi_1= \a + \vp, \ \phi_2 = \a - \vp$) 
 in  the two planes  of $R^4$ 
 which $S^3$ is embedded into\foot{The unit vector 
describing $S^2$ is related to $U_i$ by 
$
n_{i}=U^{\dagger}\sigma_{i}U,\ \ \ U=(U_{1},U_{2})\ , \  \ \ 
\vec n=(\sin 2\psi \cos
2\varphi, \sin 2\psi \sin 2\varphi, \cos 2\psi)\ . 
$
} 
\be \X_i = U_i e^{i \a}\ , \ \ \ \ \ \ \
U_1= \cos \psi \ e^{i \vp} \ , \ \ \ \ \ \ \ \ 
U_2= \sin \psi \ e^{-i \vp} \  \la{ggg}
\ee
to be constant and equal to $J$.
This is equivalent \ci{kt} 
to  gauge-fixing the 2d dual coordinate $\td \a = \J \s$. 

One possible strategy is to use the same gauge  also in 
the present case of slow strings. Then 
the spin  $J$ will again have the 
 interpretation of the length on  the spin chain side.
 The difference with the  fast
string case is that there  we had $\mathcal{J}\equiv  J/\sql $
 large so that we expanded in small
$\tilde{\lambda}=\frac{\lambda}{J^2}=\frac{1}{\mathcal{J}^2}$. For the 
slow strings we may first 
expand  in large $\lambda$, and then in  large $J$, so that
now we have $\sqrt{\lambda}\gg J$, or,  equivalently,  $\mathcal{J}\ll 1 .$
In general, quantum string corrections are
 expected to be important (modifying subleading terms in the classical action) 
but we may hope that they do not change the
 form of the leading large $\l$ term in the action. 
Proceeding as in \cite{kt}, i.e. fixing $
t=\tau, \ \  \tilde{\alpha}=\mathcal{J}\sigma$, 
we obtain the $R_t \times S^3$ string  action in the form 
\begin{equation}
I=\int dt \int_{0}^{2\pi}\frac{d\sigma}{2\pi}\ L, \quad \quad
L=J ( C_{0}-\sqrt{h} ) \ ,  
\end{equation}
\begin{equation}
h=(1+\frac{\lambda}{J^2}|D_{1}U_{i}|^2)(1-|D_{0}U_{i}|^2)+
\frac{1}{4}\frac{\lambda}{J^2}(D_{0}U_{i}^{*}D_{1}U_{i}+c.c.)^2\ , 
\label{reduced}
\end{equation}
where
$
C_{a}=-iU_{i}^{*}\partial_{a}U_{i}, \ \ 
D_aU_{i}=\del_a U_{i}-iC_aU_{i}
$.
We   expand this action  at large $\lambda$ for
 fixed $J$ and fixed derivatives
of the fields
\begin{equation}
L=-{\sqrt{\lambda}}\sqrt{|D_{1}U_{i}|^2(1-|D_{0}U_{i}|^2)+
\frac{1}{4}(D_{0}U_{i}^{*}D_{1}U_{i}+c.c.)^2}
+O(J, \lambda^{-1/2})\ . 
\end{equation}
Since for slow strings $ 1\ll J \ll \sql$  we have  ignored the first 
$J C_0$ term (which played the important role in the fast string case).
In terms of the two angles $\psi $ and $\vp$  in \rf{ggg} 
we get 
\begin{equation}
L=-{\sqrt{\lambda}}\sqrt{(\psi'^2+\varphi'^2
\sin^{2}2\psi)(1-\dot{\psi}^2-\dot{\varphi}^2 \sin^2 2\psi)+(\psi'
\dot{\psi}+\varphi' \dot{\varphi}\sin^2 2\psi)^2}+... \ . 
\end{equation}
This action does admit our basic 
circular string configuration \rf{hhh} as its solution for which 
$
\psi=\frac{\pi}{4}, \  \   \varphi=m \sigma
.$
%%%%%%%%%%%%%%%%%%%%%%%%%
We may now set $m=J$ and expand the action  near  this solution
\begin{equation} \la{kok}
\psi\rightarrow \frac{\pi}{4}+f(\tau,\sigma), \quad\quad\quad
\varphi\rightarrow J\sigma+g(\tau,\sigma)
\end{equation}
keeping all orders in the fields
but dropping  higher powers of their derivatives.
Then the $J$-dependence  can be absorbed into the
  new   spatial parameter
$$ s =  J {  \s }  $$
and we finish with $
I=\int dt \int_{0}^{2\pi J}\frac{ds}{2\pi}\ L$, where
\bea
L= \ha  \sqrt{\lambda}\bigg[&-& 2   \cos2f  - 2 g'  \cos2f  +
\cos 2f\  \dot{f}^2 -  \frac{f'^2}{ \cos 2f} \nonumber \\
&+&
 \dot{f} ( \dot{f} g' -2 f'  \dot{g})  \cos 2f +  \frac{f'^2 g' }{ \cos 2f}
 + ...\bigg]+   ....  \ , \la{pou}
\eea
where the prime now stands for the  derivative over
 $s$.

To quadratic order in  fluctuations this becomes
\begin{equation}\la{qua}
L= \ha \sqrt{\lambda}\bigg({\dot{f}^2} - {f'^2}
  + 4f^2 + ...\bigg)
\end{equation}
which represents   the unstable  mode.
Its origin is similar to the tachyonic mode appearing when expanding
the sine-Gordon
model near the maximum of the potential.

\bigskip

An alternative approach to deriving the  effective action
  is to start with  the  string action on   $R_t \times S^3$
  in a different --
  conformal -- gauge  
\be
L = - \ha \sql \bigg[  - (\del t)^2 + ( \del \a   + \cos 2 \psi\  \del  \vp)^2
+ ( \del \psi)^2   +   \sin^2 2\psi\  ( \del \vp)^2  \bigg]
\  .
\ee
For the  circular string solution \rf{hhh}  we have
$t=\kappa \tau, \ \a = { J\ov  \sql } \tau, \
\vp= J \s$,  and so for $  J\ll   \sql$ one  may
 ignore time evolution
of $\a$  and integrate out its spatial fluctuations. The resulting
Lagrangian  for $\psi$ and $\vp$  or their
fluctuations near the wrapped string solution
in  \rf{kok}   is then  (here $I
=\int d\tau \int_{0}^{2\pi}\frac{d\s}{2\pi}\ L $)
\bea
L &=& \ha \sql \bigg(  \dot \psi^2 - \psi'^2  + \dot \vp^2  -
\sin^2 2\psi\  \vp'^2 + ...\bigg) \nonumber \\
 &=&
 \ha  \sql \bigg[\dot f^2 - f'^2  + \dot g^2  -
  \cos^2 2f\  ( J +  g')^2 + ...\bigg]   \ . \la{step}
  \eea
  For large  $J$   we may replace $J +  g'\to J$ and
  thus get a weakly-coupled  combination of a sine-Gordon model for $f$
  and a free homogeneous  $g$ mode.
  %AT
  In conformal gauge the   action will then scale as $J^2 $
  but since $t=\kappa \tau \approx   J \tau $
  ($\kappa = \sqrt{ m^2 + w^2} \approx m = J$)
  the target-space energy will scale
  linearly with $J$.

  It is  useful to  rewrite the action for \rf{step}
  in terms of more  natural world-sheet
  coordinates to facilitate  comparison with
  spin chain action in the next section,
  namely, in terms of the  target-space time $t= J \tau +...$
  and $s= J { \s }$.
  The use of $s$ is natural since
  here  the length of the wound string is  large, so $J \gg 1$
  corresponds to the thermodynamic limit. Then we get
 % (ignoring for simplicity all $2 \pi$ factors)
 %AT 
  \be \la{ste}
  I
= { \sql \ov 4 \pi}    \int dt  \int_{0}^{2 \pi J}  ds \ 
 \bigg(   \dot g^2  +  \dot f^2 - f'^2    -  \cos^2 2f  + ...
 \bigg) \ ,  \ee
 and it is now obvious that the action and the energy 
 of an approximately homogeneous configurations should scale 
 linearly with large $J$.
%
% new
%

As stressed at  the beginning,
to compare to spin chain we should 
consider 
% we are interested in 
the spectrum of Hamiltonian for small fluctuations
near this slow string  state. The  Hamiltonian 
corresponding to \rf{ste} is 
\begin{eqnarray}\la{hoh}
H= \int_{0}^{2 \pi J}  ds \ \left[
\frac{\pi}{\sqrt{\lambda}}{\Pi^2_g}+
\frac{\pi}{\sqrt{\lambda}}{\Pi^2_f}
+\frac{\sqrt{\lambda}}{4\pi}(f'^2+\cos^22f)\right] \ . 
\end{eqnarray}
After a canonical transformation 
that rescales  momenta and  fields by $\sql$
 in the opposite way  we get, to quadratic  order 
 in the fluctuation field $f$ 
 (cf.  \rf{qua}) 
%implementing the fact that 
%$f$ and $g$ are fluctuations, becomes
\begin{eqnarray}
H= \ha  \int_{0}^{2 \pi J}  ds \ 
\left( {\Pi^2_g}+
{\Pi^2_f}
+ f'^2-4f^2+{\cal O}({1\ov \sqrt{\lambda}})\right) \ . 
\label{Heff2}
\end{eqnarray}
 Higher-order fluctuation terms are  suppressed 
 in the large $\l$ limit. 
 
 In the next section we shall  see that an
    effective Hamiltonian similar to \rf{Heff2}
   appears in the relevant large $\l$ limit
  on the gauge  theory spin chain side
  assuming it is described  by the  Hubbard model of  \ci{rss}.

%%%%%%%%%%%%%%%%%%%
%%%%%%%%%%%%%%%%%%%

\renewcommand{\theequation}{4.\arabic{equation}}
 \setcounter{equation}{0}

\section{An effective Hamiltonian for fluctuations near AF state 
 of  gauge theory spin chain
described  by  Hubbard model}

It has recently been shown \ci{rss}  that the Bethe equations 
diagonalizing the 
BDS spin chain \ci{bds} are identical to those diagonalizing the 
infinitely long 
Hubbard chain with the half-filled state as the ground state. 
From the standpoint of the ${\cal N}=4$ SYM theory
the most important property of
the Hubbard model is that its interactions are
short-ranged. Consequently, it can be defined on a lattice of any
length, providing {\em a possible} extension of the BDS chain to
operators of finite length.  
%Thus, the
%Hubbard model offers a possibility of taking into account some of the
%effects of wrapping interactions. 

The relation between the Hubbard model and the \adss string theory is a very
interesting question. In the event that (some modification of) it 
 represents  the correct extension of
the BDS chain to finite length operators, the Hubbard model should
also be related to the world sheet theory, 
perhaps in the same spirit 
as the Heisenberg-type  chain near the ferromagnetic end of the spectrum 
is related to  the fast string limit
of the world sheet sigma model \ci{kru}. There are important differences
however. The ground state of the half-filled Hubbard model is
anti-ferromagnetic, in the sense of possessing N\'eel order.  As was 
pointed out earlier, in the  leading 
perturbative gauge theory limit the
effective action of excitations around this state is relativistic and 
also strongly coupled. The lack of an expansion parameter
analogous to $\lambda/J^2$ raises the question of 
 how to compare this action 
to some  action derived from the  string world sheet action. 
A possibility is that  on the string  side  
 the relevant  action  may be obtained
 % (up to a change of variables)
  by integrating out all fields except those
describing the $SU(2)$ sector in the $\l \to \infty$  limit. 
Deriving such a quantum effective action 
appears to be beyond our reach at the moment.

If a version of   Hubbard model 
  does give  the correct representation for 
the gauge theory
dilatation operator 
it  would then allow to establish a contact 
with  the 
perturbative/semiclassical  (i.e. large tension or large  $\sql$) 
limit of the string world sheet theory.
% in the absence of an
%expansion parameter. 
In the large 't Hooft coupling limit, the 
effective action of small excitations around the AF 
ground state of the
Hubbard model should be compared to the classical world sheet action
expanded around the classical solution dual 
 to this ground
state. 
%
% new
%
In what follows  we shall compare 
the classical continuum limit of the 
standard Hubbard chain  with 
the effective Hamiltonian \rf{hoh} of  fluctuations around the
solution corresponding to the AF state.
%expansion of the classical string action 
 It is important to stress again  that 
this comparison is qualitatively different from that of the 
ferromagnetic case 
coherent state continuum limit and the fast string action
in \ci{kru,krt}. Rather, it
should be thought of as the comparison between the spectrum of
eigenvalues of the gauge theory dilatation operator close to some
large anomalous dimension with the eigenvalues of the effective 
fluctuations 
Hamiltonian obtained by expanding the 
string effective  action around a specific
solution. 

%
%new
%
Also, it  is 
clear that here 
%, unlike the excelent agreement at the ferromagnetic end of the
%spectrum,
 we may  not expect the  precise match between the string and spin chain
 Hamiltonians. 
As was found  in \ci{rss}, the standard 
 Hubbard model does not resolve the 
``3-loop discrepancy'', i.e. it  does not reproduce the precise  
string-theory values of subleading 
coefficients in the energy of fast-rotating  strings 
 in the large $\l$  limit; this indicates that  
 this model does not capture all the
details of the world sheet theory. 
The best we may  hope for is a 
qualitative agreement between the continuum limit of the Hubbard 
Hamiltonian  and
the slow-string effective fluctuation  Hamiltonian.

Below we  will first review the continuum limit and the bosonization of the
Hubbard model at a general filling fraction (see, 
% new ref
e.g.,  \cite{bosonization} for a recent thorough discussion).
We shall consider the  odd-length  Hubbard chain to
avoid complications related to the twist necessary for 
even lengths \cite{rss}.
We shall then  focus on  the 
 half-filling  case and  compare
the  result with the effective Hamiltonian of fluctuations 
around the slow string solution.
We shall  find a qualitative agreement.
% with its conformal gauge version \rf{ste}.
%(\ref{Lconf}) .  

\subsection{Review of continuum limit}

The Hubbard model Hamiltonian is (see, e.g., \ci{korep}) 
\begin{eqnarray}
H&=&-{\rm t}\sum_{i,\alpha}\left(
  c_{i,\alpha}^\dagger     c_{i+1,\alpha}
+c_{i+1,\alpha}^\dagger c_{i,\alpha}\right)
+U\sum_i\,c_{i\up}^\dagger c_{i\up}^{\vphantom\dagger}
                 \,c_{i\down}^\dagger c_{i\down}^{\vphantom\dagger}
\equiv H_0+H_1
\label{H0H1}
\end{eqnarray}
where $c_{i\alpha}^\dagger$ and $c_{i\alpha}^{\vphantom\dagger}$ are
creation and annihilation operators of electrons of spin
$\alpha=\{\up,\,\down\}$
at site $i$.
The relation between the two parameters $t$ and $U$ and the 't~Hooft
coupling was established in \cite{rss} by comparing the ground state
energy of the Hubbard model with the maximum energy state of the BDS
chain:
\begin{eqnarray}
{\rm t}= {\rm t}_{_{\rm RSS}}=-\frac{1}{\sqrt{2}\,{\rm g}} \ , 
~~~~~~~~
U = {\rm t}U_{_{\rm RSS}} \ , 
~~~~~~~~
U_{_{\rm RSS}}=\frac{\sqrt{2}}{\rm g} \ , 
~~~~~~~~
{\rm g}^2\equiv \frac{\lambda}{8\pi^2} \ . 
\label{rss}
\end{eqnarray}
Here ${\rm t}_{_{\rm RSS}}$ and $ U_{_{\rm RSS}}$ are the ${\rm t}$ 
and $U$ parameters used in
\ci{rss}.
%\footnote{
In the  weak gauge coupling region 
%AT
(where $U \gg {\rm t}$  and so 
the quartic term dominates over the quadratic one which is then treated as a
perturbation)
the 
effective Hamiltonian is  given by a series of the form 
\begin{eqnarray}
\sum_{k=0}^\infty {\hat A}_k\,\frac{{\rm t}^{2k}}{U^{2k-1}}
\end{eqnarray}
where ${\hat A}_k$ are operators constructed out of 
$c_\alpha^\dagger$ and $c_\alpha^{\vphantom\dagger}$.
%
%added
%
The  $k=0$ and $k=1$ terms correspond to the tree-level and
one-loop dilatation operators, respectively.
%\foot{
%ATT
%AATnew 
Let us note  that the 
normalization in \rf{rss}  is 
 different from the one usually considered:
 here the tree-level Hamiltonian contributes ${\cal
O}(1/\lambda)$ to the dimension of operators while the one-loop
Hamiltonian contributes terms   independent of the 't~Hooft
coupling.  The usual extra order-$\l$ factor 
  may be restored by rescaling both ${\rm t}$
and $U$ by $\frac{\lambda}{16\pi^2}=\frac{{\rm g}^2}{2}$.
Indeed, in \ci{rss} the energy  of the Hubbard model \rf{H0H1}
was multiplied by g$^2$  to get the anomalous dimension. 
It is more natural to define the Hamiltonian so that its 
eigenvalues are directly related to anomalous dimensions and thus, via 
AdS/CFT, to string energies. 
To implement this, 
%one could 
 here we will
  adopt the following 
``rescaled'' choice of the parameters in \rf{H0H1}:\foot{
Note that the  Bethe ansatz (Lieb-Wu) equations for the Hubbard model 
that reduce to the BDS  Bethe ansatz equations 
\ci{rss}   depend only on the  ratio $U/{\rm t}$ 
and thus are the same for the two choices.
%The possibility of such   rescaling is potentially a subtle issue.
}
\be 
 {\rm t}=\frac{{\rm g}^2}{2}{\rm t}_{_{\rm RSS}}=-\frac{\rm g}{2\sqrt{2}} \ , 
~~~~~~~~~~~~~
U={\rm t}U_{_{\rm RSS}}=-\frac{1}{2}  \ . \la{rsss}
\ee
%
% 10/01
%
The negative sign of ${\rm t}$ corrects the fact that 
the energy of the Hubbard model and the gauge theory anomalous
dimensions have opposite signs. In relation to the world sheet theory
we will choose to implement this relation by replacing ${\rm t}$ and $U$
with $|{\rm t}|$ and $|U|$ and reversing at the very end 
the sign of the time coordinate.
This will ensure  that the sigma model energies are identified
with the {\em negative} of the Hubbard model energies.
%Below we shall be interested in the opposite strong-coupling limit 
%and use the original normalization \rf{rss} 
%of \ci{rss}. 
%Note that the 
%In the
%following we will therefore use
%the normalization different  from the one in 
%\ci{rss}

%The AdS/CFT correspondence identifies the
%energies of string states with gauge theory anomalous
%dimensions. Thus, exploring the relation between the world sheet
%theory and (any representation of) the gauge theory dilatation operator
%requires that both have the same normalization.
%Let us still note that 
%} 
%}
%\begin{eqnarray}
%{\rm t}=\frac{{\rm g}^2}{2}t_{_{\rm RSS}}=-\frac{\rm g}{2\sqrt{2}} \ , 
%~~~~~~~~
%U={\rm t}U_{_{\rm RSS}}=-\frac{1}{2}  \ . 
%\label{rss_scaled}
%\end{eqnarray}
%In fact, the comparison with
%the  world sheet string theory appears to {\it require}
 %that
%we make this rescaling and thus use the choice 
% \rf{rss_scaled}.
%The AdS/CFT correspondence identifies the
%energies of string states with gauge theory anomalous
%dimensions. Thus, exploring the relation between the world sheet
%theory and (any representation of) the gauge theory dilatation operator
%requires that both have the same normalization.
%
% changed 
%
For the comparison with the classical world sheet 
string 
theory we will be interested in the
opposite limit to the one discussed in \cite{rss} -- 
 in the strong-coupling 
limit  where 
$\lambda\rightarrow\infty$.
 In this limit $|{\rm t}|\gg |U|$ and thus 
the Hubbard model as well as its continuum limit may be treated
``semiclassically'' or by expansion near the free quadratic term 
 (the quartic term in $H$ may be considered as a 
perturbation).\foot{Note that with the normalization \rf{rsss}
it is 
immediately clear  that  in the
strong-coupling limit 
the AF ground state energy should scale as 
$ {\rm t}\sim {\rm g}\sim \sql$, i.e. in the same way as  found 
by extrapolating to strong coupling \rf{ext} 
the perturbative expression \rf{bes}.
}

Our aim will be to study 
small fluctuations around the
half-filled state. The standard procedure is to construct the operators
Fourier-conjugate to $c_{j,\alpha}$ and $c_{j,\alpha}^\dagger$. The
operators creating the ground state fill up all momentum levels of 
the Fermi sea; our aim will be to find  the effective action for 
the excitations around the Fermi level, having momenta much smaller 
than the Fermi momentum $k_F$. While we are particularly interested in the
half-filled state, it is possible -- and,  in fact, 
 instructive -- to analyze 
 %in the same spirit 
the fluctuations around the minimum energy state at a general filling
fraction, i.e. for arbitrary $J_1$ and $J_2$ charges  of the $SU(2)$ sector. 
The effective Hamiltonian  obtained following this procedure could
then be compared  to  the  Hamiltonian for  fluctuations
around a classical solution dual to the minimal energy string state 
with spins  $J_1$ and $J_2$. 

The annihilation operators then are 
\begin{eqnarray}
c_{j,\alpha}=\sum_{k}e^{ikja}c_{k,\alpha}
%~~~~\longrightarrow~~~~
%c_{j,\alpha}
&\to &\sqrt{a}\left[
e^{-ik_Fja}\sum_{-k_F-\Lambda}^{-k_F+\Lambda}e^{ikja}c_{k, \alpha}+
e^{+ik_Fja}\sum_{k_F-\Lambda}^{k_F+\Lambda}e^{ikja}c_{k, \alpha} \right]\cr
                  &\equiv&
%a^{(1+\kappa)/2}
e^{-ik_Fja}L_{j,\alpha}+e^{+ik_Fja}R_{j,\alpha} 
\label{exp}
\end{eqnarray}
where $a$ is the lattice spacing and $\Lambda\ll k_F$ is a cutoff
enforcing that the fluctuations have momenta much smaller than $k_F$. 
The Fermi level of a system of length $J$ 
%(here the length of the spin chain  will be $J$) 
with $n_c$ electrons is $k_F=\pi n_c/J$; at
half-filling the number of electrons is half the number of lattice
sites and,  therefore,  we find that $2k_F a = \pi$.

Let us then 
use  the expansion (\ref{exp}) in  the 
  Hamiltonian \rf{H0H1} and take 
  the continuum limit
%\foot{It is worth  making a  comment about  the range of the space-like
%variable $x$.
%The coherent state continuum limit of the
%one-loop dilatation operator is related \cite{kru,krt} to the string
%action in the uniform gauge, which ensures that the length of the
%string is $2\pi$.  It is not immediately clear which form of  the string
%action is most appropriate for  comparison  with the continuum limit 
%considered here. For now, we will  leave the integration
%range of the space-like coordinate $x$  unspecified.} 
\be  \la{vvv} 
c_{j+1, \alpha}\simeq c_{j, \alpha}+a\partial_x c_{j,
\alpha}+\dots 
~~~~~~{\rm and}~~~~~~
\sum_j\mapsto\frac{1}{a}\int^V_0 dx~~.
\ee
%ATT
By construction, the largest value of the coordinate $x$ should be 
$V = J a$. One possible  choice  used 
 in the near-ferromagnetic  ground state
case \ci{krt} is  $a= { 2 \pi \ov J}, \ V= 2 \pi$; in that  case 
the  world-sheet coordinate  had $J$-independent length while 
the  $J$-factors combined in the scaling limit with $\sql$. 
Here  we shall use instead $a=1, \ V=J$; this is natural since 
in the thermodynamic limit  $J \gg 1$ 
all extensive quantities describing near-AF states should 
scale linearly with $J$. 
The coordinate $x$ will then  be directly related to $s$ in \rf{ste}
up to $2 \pi$ factor. 
For generality we shall keep the  dependence on the 
lattice spacing $a$ explicit in what follows. 

%the world-sheet coordinate should have 

Plugging (\ref{exp}) into the quadratic and quartic terms of
(\ref{H0H1})  leads to:

1) the quadratic Hamiltonian:
\begin{eqnarray}
%&&\!\!\!\!\!\!\!\!\!\!\!
%\frac{H_0}{a^\kappa} 
H_0\!\!\!&=&\!\!\!
-|{\rm t}|\sum_{j, \alpha}\left[\cos k_Fa\,
(L^\dagger_{j,\alpha}L^{\vphantom\dagger}_{j,\alpha}
+
R^\dagger_{j,\alpha}R^{\vphantom\dagger}_{j,\alpha})
+2ia
%a^{2}t\sum_{j, \alpha}
\sin k_Fa\,
(R^\dagger_{j,\alpha}\partial_xR^{\vphantom\dagger}_{j,\alpha}
-
L^\dagger_{j,\alpha}L^{\vphantom\dagger}_{j,\alpha})
+\dots\right]\nonumber \\
&\sim&
%\!\!\!\!
- \frac{|{\rm t}|}{a}\cos k_F a\,\sum_\alpha\int dx
(L^\dagger_{\alpha}L^{\vphantom\dagger}_{\alpha}
+
R^\dagger_{\alpha}R^{\vphantom\dagger}_{\alpha}) \nonumber \\
&+&
2\,|{\rm t}|\,\sin k_F a\sum_{\alpha}\,\int\,dx\,
(L^\dagger_{\alpha}\,i\partial_xL^{\vphantom\dagger}_{\alpha}
-
R^\dagger_{j,\alpha}\,i\partial_xR^{\vphantom\dagger}_{j,\alpha})
\label{H0}
\end{eqnarray}
In writing the first line in the 
equation above we discarded summands proportional to 
$e^{\pm 2ik_Fja}$; 
the reason is that, upon Fourier transforming $L$ and $R$,
the sum over $j$ vanishes due to the assumption that the momenta of
the excitations are much smaller than the Fermi momentum $k_F$.
%We will shortly see that the first term in \rf{H0}  is irrelevant.

2) the quartic  Hamiltonian:
%As before, discarding summands proportional to 
%$e^{\pm 2ik_Fja}$, the quartic Hamiltonian becomes:
\begin{eqnarray}
H_1
&=&
%a^{2+2\kappa}
|U|\sum_{j}\left[
(:\!L_{j\up}^\dagger L_{j\up}^{\vphantom\dagger}\!:
+:\!R_{j\up}^\dagger R_{j\up}^{\vphantom\dagger}\!:)
(:\!L_{j\down}^\dagger L_{j\down}^{\vphantom\dagger}\!:
+:\!R_{j\down}^\dagger R_{j\down}^{\vphantom\dagger}\!:)
+ (L_{j\up}^\dagger R_{j\up}^{\vphantom\dagger}
R_{j\down}^\dagger L_{j\down}^{\vphantom\dagger}
+
h.c.
%R_{j\up}^\dagger L_{j\up}^{\vphantom\dagger}
%L_{j\down}^\dagger R_{j\down}^{\vphantom\dagger}
)
\right]\cr
&+&
%a^{2+2\kappa}
|U|\sum_j\left[
e^{+4ik_Fja}L_{j\up}^\dagger R_{j\up}^{\vphantom\dagger}
           L_{j\down}^\dagger R_{j\down}^{\vphantom\dagger}
+
e^{-4ik_Fja}R_{j\up}^\dagger L_{j\up}^{\vphantom\dagger}
           R_{j\down}^\dagger L_{j\down}^{\vphantom\dagger}
\right] \la{sec} 
\end{eqnarray}
We have again  discarded  summands proportional to 
$e^{\pm 2ik_Fja}$. 
Away from half-filling the second line  in \rf{sec}  is irrelevant. 
At half-filling we have $e^{\pm 4\pi i a}=1$ which 
leads to the survival of the
second line in \rf{sec} or in the effective action. 
Introducing the parameter $\zeta$
which vanishes away from half-filling and equals unity at half-filling,
it follows that the continuum limit of the quartic part of the Hubbard 
Hamiltonian expanded around the Fermi levels is
\begin{eqnarray}
H_1&=&
%a^{1+2\kappa}
\frac{|U|}{a}\int\,dx\,\left[
(:\!L_{\up}^\dagger L_{\up}^{\vphantom\dagger}\!:
+:\!R_{\up}^\dagger R_{\up}^{\vphantom\dagger}\!:)
(:\!L_{\down}^\dagger L_{\down}^{\vphantom\dagger}\!:
+:\!R_{\down}^\dagger R_{\down}^{\vphantom\dagger}\!:)
+
L_{\up}^\dagger R_{\up}^{\vphantom\dagger}
R_{\down}^\dagger L_{\down}^{\vphantom\dagger}
+
R_{\up}^\dagger L_{\up}^{\vphantom\dagger}
L_{\down}^\dagger R_{\down}^{\vphantom\dagger}
\right]\cr
&+&
\zeta\,
%a^{1+2\kappa}
\frac{|U|}{a}\int\,dx\,\left(
L_{\up}^\dagger R_{\up}^{\vphantom\dagger}
L_{\down}^\dagger R_{\down}^{\vphantom\dagger}
+
R_{\up}^\dagger L_{\up}^{\vphantom\dagger}
R_{\down}^\dagger L_{\down}^{\vphantom\dagger}
\right)
\label{H1}
\end{eqnarray}

%
%new
%
To summarize, the equations (\ref{H0}) and (\ref{H1}) represent the 
Hamiltonian of the fluctuations around the half-filled state
($\zeta=1$) and the state at generic filling ($\zeta=0$) of the
Hubbard model. 
We would like to compare the large $\l$ limit
(or linearized) spectrum of this fluctuation  Hamiltonian 
 to the spectrum of the string  Hamiltonian 
\rf{hoh} or (\ref{Heff2}).
The first step is then to bosonize (\ref{H0}),(\ref{H1}).

%ATT
%An important question  is about  the dependence on the lattice spacing 
%$a$.
% and its relation to the rescaled RSS identifications
%(\ref{rss_scaled}). 
%If we choose $a=1$  in \rf{vvv} 
% then in the large $J$ limit the continuum theory is defined
%on the  circle of infinite length. 
%If instead  we choose $a\sim { 1 \ov  J} $  (so that 
%is  $x$  spanning a finite-radius
%circle)  
%then at half-filling  the  last quadratic  term in \rf{H0} 
%will have  coefficient  $\rt $  while the quartic  term \rf{H1}
%will scale as  $ U J$. 
%A natural scaling for the near AF-state energies would be 
%linear growth with $J$ as  found also on the string side. 
%For generality, we  will keep the lattice spacing dependence 
%explicit in the following. 
%
% added
%
%In the end we will choose the lattice
%spacing which yields the closest match with the slow string action.

%Ageneral, it is possible to choose a scaling for the parameters
%A${\rm t}$ and $U$ such that the resulting continuum-limit 
%A action is finite.\footnote{
%ANote that this would require that $\rm t$ is finite while $U$ scales to
%Azero, which is in line with the theory been  in the perturbative regime.
%A}  
%AIn the present  case 
%Athe equation (\ref{rss_scaled}) implies that 
%A$U$ is a constant, so  that it is
%Anot possible to scale it in such a way that all terms in the action are
%Afinite.
%ABecause of this issue
%Awe will keep the lattice spacing dependence explicit in the following.  

\subsection{Bosonization of the continuum-limit Hamiltonian}

There are three ways to relate the above fermionic Hamiltonian 
 to a bosonic 
theory. One  -- which  we will follow  here -- is to
directly bosonize the Hamiltonians  (\ref{H0}) and (\ref{H1}).  Another 
 is to express the continuum limit of $H$   
in terms of the $SU(2)\times SU(2)$ currents \cite{korep,tsvel};   the 
third possibility is to use a mean field approximation \cite{frad}.
The latter two approaches yield a direct sum of the 
conformal  $SU(2)$ level one  WZW model 
and a massive $U(1)$ Thirring model. 
This  representation of the scaling-limit theory  is not bosonic and
thus is not suitable for comparison with the 
slow-string actions (\ref{pou})  or (\ref{ste}). However,
 the WZW model is equivalent to a 
compact boson at self-dual radius, while the Thirring model is
equivalent to a sine-Gordon model. 
In the end, all three approaches are equivalent, leading to the results obtained by directly
bosonizing   (\ref{H0}) and (\ref{H1}) as  discussed below.

Using the rather standard bosonization formulae
($\gamma$ is for the time being an arbitrary constant) 
\begin{eqnarray}
&&:\!L_\alpha^\dagger\, L_\alpha^{\vphantom\dagger}\!:=
{\gamma^2}\partial\Phi_{L\alpha}\ , 
~~
L_\alpha={\gamma}e^{-i \,\Phi_{L\alpha}}\ ,
~~
\Phi_{L\alpha}(z)\Phi_{L\alpha'}(0)=-{\delta_{\alpha\alpha'}}
\ln z\ , \ \ \ \ \cr
&&:\!R_\alpha^\dagger R_\alpha^{\vphantom\dagger}\!:=
{\gamma^2}\partial\Phi_{R\alpha}\ ,
~~
R_\alpha={\gamma}e^{~ i \,\Phi_{R\alpha}}\ ,
~~
\Phi_{R\alpha}(z)\Phi_{R\alpha'}(0)=-{\delta_{\alpha\alpha'}}
\ln z\ , 
\label{bosonization}
\end{eqnarray}
translated into the Hamiltonian formalism\foot{The bosonization formulae 
used here apply to a theory defined on
the plane $\pRe^2$, with $z=re^{i \theta }$. For the purpose of
comparison with string theory we need to pass to $\pRe\times S^1$,
where the time-like direction is related to $r$ as 
$r=e^{\hat \tau_E}= e^{i\hat \tau }$.
%AAT
Note that as usual translations (but not rescalings) of the 
(euclidean)   time 
coordinate 
correspond to dilatations
on  $\pRe^2$ (under which the operators 
in \rf{bosonization} have appropriate 2d quantum dimensions). 
}
%but rescaling of $\tau$  does not correspond to a 
% We see thyerefore that translation
%in time (or, more generally, redefinitions of the time coordinate) 
%are unrelated to conformal transformations.}
%(essentially set $t=0$), 
we are quickly led to the following bosonic Hamiltonian 
\begin{eqnarray}
%\frac{H}{a^{1+2\kappa}}
H&=&2\gamma^2 |{\rm t}|\sum_{\alpha=\up,\down}\,\int\,dx\,
\left[(\partial_x\Phi_{L,\alpha})^2
+
 (\partial_x\Phi_{R,\alpha})^2\right]\cr
&+&\!\!\!\!
\gamma^4\frac{|U|}{a}\int dx\,\bigg[
(\partial_x\Phi_{L{\up}}+\partial_x\Phi_{R{\up}})
(\partial_x\Phi_{L{\down}}+\partial_x\Phi_{R{\down}})
%\cr  
%&&~~~~~~~~~~ 
+ 2\cos
(\Phi_{L\up}-\Phi_{L\down} -\Phi_{R\down}+\Phi_{R\up})
\bigg]\cr
&+&\!\!\!\zeta\,\gamma^4
\frac{|U|}{a}\int dx\,\ 2 \cos (\Phi_{L\up}+\Phi_{L\down}
+\Phi_{R\up}+\Phi_{R\down}) \ . 
\end{eqnarray}
The commutation relations of the original creation
and annihilation operators imply certain commutation relations between
the fields
\begin{eqnarray}
\phi_\alpha \equiv  \Phi_{L\alpha}+\Phi_{R\alpha}\  , 
~~~~~~~~ \ \ \ \ \ \ \ 
\theta_\alpha  \equiv   \Phi_{L\alpha}-\Phi_{R\alpha}~~.
\label{conjugate_coord}
\end{eqnarray}
In particular, it turns out that $\partial_x\theta_\alpha$ 
can be interpreted as the momentum conjugate to $\phi_\alpha$,
implying that the Hamiltonian simplifies to
\begin{eqnarray}
H&=&\gamma^2\,
|{\rm t}|\,\int\,dx\,
\left[\big(\Pi_\up^2+\Pi_\down^2\big)
+(\partial_x\phi_{\down})^2
+(\partial_x\phi_{\up})^2\right]\\
&+&\gamma^4
\frac{|U|}{a}\int\,dx\,\left[
\,\partial_x\phi_{\up} \partial_x\phi_{\down}
+2\cos(\phi_{\up}-\phi_{\down})
+
2\zeta\cos(\phi_{\up}+\phi_{\down})
\right]
\nonumber
\end{eqnarray}
Furthermore, this  Hamiltonian can be rewritten as a sum of two decoupled 
Hamiltonians by introducing 
\begin{eqnarray}
\phi_c=\frac{1}{\sqrt{2}}(\phi_\up+\phi_\down)\ , 
~~~~~~~~~~~~~
\phi_s=\frac{1}{\sqrt{2}}(\phi_\up-\phi_\down) \ . 
\end{eqnarray}
We then get:\footnote{The $c,s$ notation is used to emphasize 
 the important well-known
fact that the excitations rearranging the  spin and the 
 charge distributions are
decoupled in the Hubbard model \ci{korep}.}
\begin{eqnarray}
H&=&H_s+H_c
\label{Hfinal}
\end{eqnarray}
with\foot{
%
% 10/01
%
We recall that, since we replaced ${\rm t}$ and $U$ by $|{\rm t}|$ and
$|U|$, to match the energy of the bosonized continuum Hubbard model
with that of a world-sheet  theory we need to reverse the sign of the
time coordinate. This transformation has no effect at the level of the
Hamiltonian or the Lagrangian, since they have an even number of 
time derivatives.
%For comparison  with bosonic field theory it is useful to change the
%overall sign of the Hamiltonian to make it positive; we do this by replacing 
%$\rt$ by $|\rt|$.
}
\begin{eqnarray}
\frac{1}{\gamma^2 |{\rm t}|}H_s&=&
\int\,dx\,
\left[\Pi_s^2+ (1-\frac{
\gamma^2 U}{4 a {\rm t}}) (\partial_x\phi_{s})^2
+2\hphantom{\zeta}\frac{
\gamma^2 U}{a\,{\rm t}}\cos (\sqrt{2} \phi_s)\right]
\label{Hs}\\
\frac{1}{\gamma^2 |{\rm t}|
}H_c&=&\int\,dx\,
\left[\Pi_c^2+  (1+\frac{
\gamma^2 U}{4 a {\rm t}})  
(\partial_x\phi_{c})^2
+2\zeta\, \frac{\gamma^2 U}{ a\,{\rm t}}\cos (\sqrt{2}\phi_{c})\right]
\label{Hc}
\end{eqnarray}
Thus apparently we end up with two sine-Gordon theories. 

As is well known, 
the continuum limit of the
Heisenberg ${\rm XXX}_{1/2}$ chain near  the anti-ferromagnetic  state 
is described by a   relativistic 2-d  theory  \ci{frad,tsvel}. 
The excitations of the Heisenberg chain
span only a subset of the excitations of the Hubbard model, namely 
(up to duality transformations), 
 only those in which all sites are either empty or
doubly-occupied. 
Taken separately and after  appropriate redefinitions
of the space-like coordinate, each of the two Hamiltonians \rf{Hs} and \rf{Hc} 
 can be interpreted as describing a relativistic theory. 
 However, if they are  combined  together, 
the relativistic theory interpretation 
is  not possible because the  speeds of light
for the two types of decoupled  excitations are 
{\it different}:
\begin{eqnarray}
v_s=\sqrt{1-\frac{
\gamma^2 U}{4 a\,{\rm t}}}=\sqrt{1-\frac{\gamma^2 }{2\sqrt{2}\,a\,{\rm g}}}
\ ,   ~~~~~~~~
v_c=\sqrt{1+\frac{
\gamma^2 U}{4 a\,{\rm t}}}=\sqrt{1+\frac{\gamma^2}{2\sqrt{2}\,a\,{\rm g}}}
~~.
\label{sofl}
\end{eqnarray}
It is important to emphasize that in constructing this continuum limit
we have assumed that the Hubbard coupling constant $U$ is small
compared to $\rm t$, i.e. $\rm g$ should be large enough. 
 This is
reflected in the above expressions \rf{sofl}
in  that the positivity of the
Hamiltonian \rf{Hfinal} 
implies 
 that we are not allowed to take the 't Hooft coupling  or ${\rm g}^2$ 
 to be
arbitrarily small. In other words, as expected from the analysis of
the discrete Hamiltonian, recovering the  perturbative region of the 
gauge theory  dilatation operator 
 requires quantum treatment  of the 
Hubbard  model of \ci{rss}. 

%
%new
%
The bosonized Hamiltonians (\ref{Hs}) and (\ref{Hc}) are,  however,  not
the end of the story. Their sum, while  
looking similar
to the effective Hamiltonian \rf{hoh} of fluctuations around the slow-string
solution, is qualitatively different from \rf{hoh}: 
% in the conformal gauge \rf{ste}, 
 both  
fields appear to
be interacting at half-filling ($\zeta=1$), while one of the fields of the
slow-string action \rf{ste} or in  \rf{hoh} is free in the large $J$ limit. 

To find a way to match (\ref{Hfinal}) and \rf{hoh}, (\ref{Heff2})
\footnote{The most naive suggestion --
to depart from the half-filling -- is of course not an option, since the slow
string action (\ref{ste}) was derived by assuming that 
we are expanding the string action
around the solution dual to the half-filled state.} 
let us  analyze  (\ref{Hs}) and (\ref{Hc}) separately. 
% treating them as relativistic theories.
 Through a canonical transformation the
 speed of light factor  can be moved into  the argument of the cosine
 potential.
Then, in the free theory approximation (which is valid as $\l$ is assumed to be
large), the dimensions of the operators
representing the potential terms 
\begin{eqnarray}
{\cal O}_{s,c}=\cos
\frac{\sqrt{2}\phi_{s,c}}
{\left(1\mp\frac{\gamma}{2\sqrt{2}a{\rm g}}
\right)^{1/4}}
\end{eqnarray}
are
\begin{eqnarray}
d_{_{{\cal O}_{s,c}}}=\frac{2}{
\sqrt{1\mp\frac{\gamma^2}{2\sqrt{2}a{\rm g}}}}~~ \ .  
\end{eqnarray} 
This means that the  interaction term
is  an {\it irrelevant operator} in $H_s$ but relevant one in
$H_c$ (with $\zeta=1$). 
From the standpoint of the world sheet infrared physics we
can therefore replace $H_s$ by a free (gapless) 
Hamiltonian. 

%
%new
%
As a result, the effective Hamiltonian for small fluctuations around
the half-filled state of the Hubbard model is
\begin{eqnarray}
H&=&\gamma^2|{\rm t}|
\int_0^{aJ}dx 
\left[~~
\Pi_c^2+  (1+\frac{
\gamma^2 U}{4 a {\rm t}})  
(\partial_x\phi_{c})^2
+2\, \frac{\gamma^2 U}{ a {\rm t}}\cos (\sqrt{2}\phi_{c})
\right.\cr
&&\left.~~~~~~~~~~~~~~~~~~~~~~
+\Pi_s^2+ (1-\frac{
\gamma^2 U}{4 a {\rm t}}) (\partial_x\phi_{s})^2
~~~\right]
\end{eqnarray}
%Here $\hat \tau$ is the time coordinate  conjugate to the Hubbard
%Hamiltonian,  i.e.  the corresponding energies should be equal to
%anomalous dimensions of gauge-theory operators.
Next, let us choose 
 the free parameter $\gamma$ such that the second  velocity is zero,  
 $v_s=0$, i.e. 
\be\la{gag}
 \gamma^2= {4a{\rm t}\over U} \ .  \ee 
Introducing the rescaled  fields 
\be { f}=\frac{\phi_c}{2\sqrt{2}} \ , \ \ \ \ \ \ \ \ \ \ \ \ 
{ g}= \frac{\phi_s}{2\sqrt{2}} \ ,   \la{fg} \ee
  we are then led  to 
  %(ignoring  a constant shift)
\begin{eqnarray}\la{hih}
H = \gamma^2 |{\rm t}|  \int_0^{aJ}dx 
\left(~~\Pi_g^2+\Pi_f^2+16\,(\partial_xf)^2
+16\,\cos^2 2f\right) \ . 
\end{eqnarray}
The 
identifications (\ref{rss}) combined with the 
choice of $\gamma$ in \rf{gag}
lead to 
\be \gamma^2|{\rm t}|
%= {\rm g}^2 a =    { \l \ov 8 \pi^2} a 
=   { 4 a |{\rm t}|^2 \ov  |U| } =  a{\rm g}^2 
\ . \ee
Moreover, choosing, as discussed above, 
 the lattice spacing to be $a=1$,
 we conclude that 
%
% 10/01
%
%the factor in front of 
the Hamiltonian \rf{hih}
% is constant. Then \rf{hih} 
has essentially the same 
 structure as  \rf{hoh} apart from order $\l$ factors.

 %Canonically rescaling  the momenta and the fields
 % and 
%then linearizing $H$
We then find  the following effective Hamiltonian for the linearized
fluctuations around the half-filled state:
\begin{eqnarray}
H=  {\rm g}^2  \int_0^{J}dx \bigg( 
 \Pi_g^2+\Pi_f^2+ 
 16\left[(\partial_xf)^2-4 f^2~\right]+\dots\bigg) \ .
\label{final}
\end{eqnarray}
The relative coefficients here can be adjusted further by 
canonically rescaling  the momenta and the fields.

There are quite obvious similarities between this Hamiltonian \rf{final}  and the 
Hamiltonian of the fluctuations around the slow string solution
(\ref{Heff2}): both describe a massive and a massless field and the ratio
between the mass and the mode number of the massive field is also the 
same in the two Hamiltonians. 
As already mentioned, the  target-space time coordinate 
$t$ in \rf{ste} should be identified 
%
% 10/01
%
(due to our choice of sign for
the couplings of the Hubbard model) 
with 
the sign-reversed time coordinate conjugate to Hubbard's Hamiltonian 
%$\hat \tau$ 
to ensure that 
 the string energies   match anomalous dimensions
on the spin chain side.
%any further rescaling of $\hat \tau$  would spoil ths identification. 
The spatial coordinates $x$ in \rf{final} and $s$ in \rf{ste}
are essentially the same,  modulo the 
$2 \pi$ factor.  

%ME%
The    coefficients in the two Hamiltonians, however,  
%coefficient between the momentum and the mass 
appear  to
be different:  $H$ in \rf{final} has an extra overall factor of 
$ {\rm g}^2=\frac{\l }{8 \pi^2 }$ while it is absent in \rf{Heff2}.\foot{It is
curious that this  is the same rescaling  coefficient that we needed to 
introduce to make the Hubbard model energies match the spin chain and thus string theory
ones;
had we used the un-rescaled  choice of Hubbard model  parameters 
\rf{rss} we would not get that overall factor in \rf{final}.
At the moment we do not understand if this is a coincidence or 
an indication that we have missed a compensating $ {\rm g}^2$
factor at some other step.}
One may say this  is hardly unexpected,  given, in particular, 
 the
$1\ov \pi^2$  mismatch between  the AF ground state energy of Hubbard
model  \rf{ext}  and the slow-string energy in \rf{cla}, 
%
% 10/01
%
to leading
order in $\sqrt{\lambda}$.

To  appreciate additional 
subtleties that one may  need to  overcome 
on the way to better understanding the correspondence 
between the near-AF state spin chain described by Hubbard 
model  and the slow-string limit on the string side 
it is instructive to   consider
the continuum limit for fluctuations 
 around the minimal energy state at some arbitrary
filling fraction. On the one hand,  this limit should
correspond to an  effective Hamiltonian for fluctuation around the
semiclassical solution dual to the $(J_1,J_2)$ operator of maximal
anomalous dimension.\foot{For $J_i\ll \sqrt{\lambda}$ and before
expanding to quadratic order, this should 
be a ``slow string'' Hamiltonian, of the same type as 
(\ref{pou}) or (\ref{ste}).}  
It is reasonable to expect that at least one of the two 
fields of appearing in the slow-string effective Hamiltonian will be 
interacting in general and massive at the quadratic level. On the other
hand,  as we have seen earlier in this section, 
  away from the half-filling we are to set 
  $\zeta=0$. Then  the interaction term in the
Hamiltonian (\ref{Hc}) vanishes and the continuum limit as constructed
above yields a free theory.  It appears,  therefore, that the
 qualitative 
agreement that we  have described above 
is restricted to the half-filled Hubbard model.

It would be 
interesting to understand if considering an effective action 
including  other
degrees of freedom would yield a better match away from the 
half-filling or, if possible, find a modification of the Hubbard
Hamiltonian which does not affect the weak 't~Hooft coupling limit,
preserves integrability and accounts for the additional interaction in
the strong 't~Hooft coupling limit.

% new ref
There is an intriguing similarity between the discontinuous behaviour
of the effective Hamiltonian of the Hubbard model 
\footnote{This discontinuity is an example \cite{Hubb1} of the
so-called  Mott-Hubbard
metal-insulator phase transition \cite{Hubb1,MIT}.} 
and that of
the $SU(2)$ sector of gauge theory at strong coupling, i.e. 
 from the point of view of the world sheet theory.
As it has been  discussed in \cite{minah}, while the
excitations around the $J_1\ne J_2$ states mix with  other 
``non-$SU(2)$''
world sheet excitations, they could be decoupled if $J_1=J_2$. It is 
tempting to conjecture that the differences between the Hubbard model
and the slow string effective Hamiltonian away from half-filling 
can be corrected by additional interaction terms in the Hubbard
Hamiltonian which account for 
mixing with other gauge theory operators.

\renewcommand{\theequation}{5.\arabic{equation}}
 \setcounter{equation}{0}

\section{Some  ``slow'' string solutions with spins in $AdS_5$ and $S^5$ 
  } 
  
The general  case of noncompact sectors is different: 
there is apparently no  bound on the quantum string energy. 
One may relate this  to the fact 
that the string wrapped on a circle in $S^3$ part of $AdS_5$
can not be static  and in any case  can have any radius (and thus any energy). 
It is still useful  to study ``slow-string'' limits 
of solutions that carry one spin  ($S)$  in $AdS_5$  and one spin ($J$) 
in $S^5$ as they may have some interpretation 
in the $SL(2)$ sector of gauge theory. 
In particular, we shall find that 
 there is again a case  in which  the classical string energy 
scales as $E \sim \sql  J + ... $. 

Below we shall discuss limits of the circular $(S,J)$ solution 
of \ci{art} and also consider  a  ``flat-space like'' solution 
which may be viewed as a special case  of the 
more general $(S,J_1,J_2)$  circular  solution  in \ci{art}.

\subsection{Circular  solution in $AdS_3 \times S^1$ }

Let us review the form of the solution of \ci{art}
describing a string  which has a rigid  circle form in $AdS_5$ 
and in $S^5$ and each circle rotates ``along itself''. 
In terms of complex combinations of global embedding coordinates 
($\Y_i$  in $AdS_5$ and $\X_i$ in $S^5$) 
one has: 
\bea
  \Y_{0}=r_{0} \; e^{i\kappa \tau}\
 , \quad \Y_{1}=r_{1} \;e^{i\w \tau+ik\sigma}\ ,
  \quad \X_1=e^{iw\tau-im\sigma}\ , \ \  \ \ \ \Y_2=\X_2= \X_3=0\ , 
 \label{solold}
 \eea
 \begin{eqnarray} \la{rad}
 \r_0 \equiv \cosh \rho_{0}\ , \quad\quad \ \ \ \
 \r_1\equiv \sinh \rho_{0}\ , \ \ \ \ \ \ \ \ \ \ \ \ \
  \r_0^2 - \r_1^2 =1 \ ,
 \end{eqnarray}
where $\rho_{0}=$const,  and $k$ and $m$ are positive integer 
  winding numbers. The
charges are
\begin{equation}
E\equiv \sql \E=\sqrt{\lambda}\kappa \r_0^2,\  \quad S\equiv \sql \S = 
\sqrt{\lambda}\w \r_1^2,\ 
\quad J\equiv \sql \J= \sqrt{\lambda}w \ . 
\end{equation}
The equations of motion imply
$
\w^2=k^2+\kappa^2
$ and 
 the conformal gauge constraints give 
\begin{equation}
2 \kappa \mathcal{E}-\kappa^2=2
\mathcal{S}\sqrt{k^2+\kappa^2}+\mathcal{J}^2+m^2
\end{equation}
\begin{equation}
k\mathcal{S}   =m \mathcal{J}\ . 
\end{equation}
The energy is thus a function of the three  parameters, e.g., 
 $E=\sql \E( { J\ov \sql},  { S\ov \sql}, m)$. 

Let  consider the special case of  
$$ S=J, \ \ \ \ \  {i.e.} \ \ \ \ \ \    k=m  $$  
Then  the independent parameters are
$\mathcal{J}=w$ and $m$  and 
\begin{equation}
2\kappa \mathcal{E}=(w+\sqrt{m^2+\kappa^2})^2\ , \quad\quad
\frac{\mathcal{E}}{\kappa}=\frac{w+\sqrt{m^2+\kappa^2}}{\sqrt{m^2+\kappa^2}}
\end{equation}
Solving for $\kappa$ we obtain
\begin{equation}\la{sii}
\kappa_{\pm}=\sqrt{\frac{w^2}{2}+m^2\pm
\frac{w}{2}\sqrt{w^2+8m^2}}
\end{equation}
Note that the minus sign solution can exist only if $m\geq
w$. The energy is 
\begin{equation}\la{si}
\mathcal{E}_{\pm}=\frac{\bigg(w+\sqrt{m^2+\frac{1}{2}(w^2+2m^2\pm
w\sqrt{w^2+8m^2})}\bigg)^2}{\sqrt{2w^2+4m^2\pm
2w\sqrt{w^2+8m^2}}}
\end{equation}
Like in the  case of the $S^5$ solution with $J_1=J_2$
 energy of the $SL(2)$ sector  solution with $S=J$ thus
  has an explicit analytic 
form. 

Small fluctuations  near this solution were discussed in
\cite{ptt}. There are 4 real massive fields from $S^{5}$ with mass
$\sqrt{w^2-m^2}$, i.e. $\omega_n=\sqrt{n^2+w^2-m^2}$. This 
frequency is real if
$
n^2+w^2-m^2\geq0   . 
$
From $AdS_{5}$ there are also 2 free massive real fields with mass
$\k$.
Remaining   fluctuations 
are coupled and the corresponding characteristic equation is 
\begin{equation}  \la{sta}
(\omega_n^2-n^2)^2+4 r_{1}^2 \kappa^2
\omega_n^2-4(1+r_{1}^2)(\sqrt{m^2+\kappa^2}\omega_n +n m)^2=0
\end{equation}
where $r_{1}^2=w/\sqrt{m^2+\kappa^2}$. The 
 stability condition is the
reality of the solutions of  this quartic equation.

In the   standard semiclassical expansion one  assumes
that $m, w$ are fixed while $\lambda$ is large.
As in the above $S^5$ solution 
 case we  can now consider  
particular limits of the parameters:\foot{The case when $k=m=0$ is again
the BPS one, $E-S=J$, when string world surfaces degenerates 
to a massless geodesic. Here the string trajectory  is a massive geodesic 
in $AdS_5$  and a big circle 
 in $S^5$;  to make canonical identification between the 
string and gauge states/energies 
  one is to  apply an $AdS_3$ transformation to transform the $AdS_5$ geodesic to
  rest frame, $t=\tau$.
}

% (i)\  $m=0$: \ \ \ \ this is ``BPS'' case   when $E=2J.$

(i)\ $w\gg m$:\ \ \  this is the ``fast string'' case \cite{art}.
Only the solution with plus sign is possible.
 The energy has a 
regular expansion in ${m^2 \ov w^2} = m^2 \tl= \frac{m^2 \lambda}{J^2}$
\begin{equation}
E=J\bigg(2+\frac{m^2 \lambda}{J^2}-\frac{5}{4}\frac{m^4
\lambda^2}{J^4}+...\bigg) \ . 
\end{equation}
 As was shown in \cite{art},  this
solution is stable for large $w$.

 (ii) $w=m$: \ \ \  this  a ``flat-space'' type
solution. We get 
\begin{equation}
E=\frac{3\sqrt{3}}{2} J \ . 
\end{equation}
Similar  ``flat-space''-type solution will be discussed in
the next subsection. As follows from \rf{sta}, 
this solution may be unstable for certain values of $w$. 

%The quartic equation becomes
%\begin{equation}
%(\omega^2-n^2)^2+6 w^2 n^2-6 w^2 (n+2 \omega)^2=0
%\end{equation}
%An analysis of the solution of this equation reveals that there
%are domains in ($n,w$) in which all solutions are real but also
%domains where some solutions are imaginary.

 (iii)\ $w\ll m$: \ \ \ this is a slow-moving  string: the $\tau$
part of the solution is much smaller then the winding $\sigma$
part.\foot{A different large winding limit of the $S \not= J$ solution was considered
in \ci{szz}.}
 There are two possible
  cases for the two signs in \rf{sii}.
   We will concentrate only on the solution with plus
sign, as the other one  can be treated similarly. 
Here we can
expand $E= \sql m\  F({  w \ov m}) $ in  $w/m$, i.e.   
\begin{equation}
E=\sqrt{\lambda}m+
\sqrt{2}J+\frac{J^2}{4m}\frac{1}{\sqrt{\lambda}}-
\frac{J^3}{8\sqrt{2}m^2}\frac{1}{\lambda}+...\ . 
\end{equation}
Here  $m\ll
J=\sqrt{\lambda}\ w$ since $\lambda$ is taken large first.
 As in the $SU(2)$ case this  solution always has unstable modes: 
  the  condition of reality of characteristic frequencies of $S^5$ fluctuations 
  is 
$n\geq m$. 
The fluctuations in other directions are non-tachyonic:  expanding the 
solutions of the quartic
equation \rf{sta} at large $m$ we find that all frequencies are real in
this limit.

 Like in the $SU(2)$ case,  we can then increase $m$ further, 
 but here one is not expecting the upper bound on the 
 string energy so there should 
 be no obvious choice for the maximal $m$.\foot{ 
 The absence of an upper bound on the string energy is consistent with 
 gauge-theory expectations in the $SL(2)$ sector: 
 for fixed $J$, i.e. fixed    length of the chain,
    the spin-chain energy can be arbitrarily 
large  because the spin  $S$ can be arbitrarily  large.
We are grateful to K. Zarembo for this remark.}
 Still, let us   formally consider  again the case of $m=J$ 
 (which corresponds to $w = { m \ov \sql} \to 0$ in the 
 large $\l$ limit).
  Although it is not
 clear which  state on the gauge  spin chain side should correspond to the 
 $m=J$ string state, let us discuss this case  by
 analogy with the  $SU(2)$ case.  Setting $m=J$ in   \rf{si} 
 one
 obtains (we choose plus sign) 
\begin{equation}
{E}= {J}\frac{\bigg(2+\sqrt{2+8\lambda +
2\sqrt{1+8\lambda}}\bigg)^2}{4\sqrt{2+4\lambda +
2\sqrt{1+8\lambda}}}  \ , \label{energy}
\end{equation}
i.e.  at 
large $\lambda$ 
\begin{equation}
E=\sqrt{\lambda}J\bigg(1+\frac{\sqrt{2}}{\sqrt{\lambda}}+\frac{1}{4\lambda}
 + ...\bigg)\ . 
\end{equation}
As in the
$SU(2)$ case we computed the 1-loop correction to the energy for the
 case of $m=J$ and found that it depends linearly on large  $J$
 (details are given in 
 Appendix).  This suggests that  in general for large $J$ one should have 
 $E=f(\lambda) J $;    this relation  may 
 then be extrapolated to weak coupling and should
 correspond to the anomalous dimension of  a particular  state 
 in the spectrum of the  $SL(2)$ spin chain.

\subsection{``Flat-space'' type $(S,J)$   solution in $AdS_{3}\times S^{2}$}

Let us now consider  another example 
of an $(S,J)$ 
``flat-space''  solution which may be viewed as a special case 
of the rational $(S,J_1,J_2)$ solution in \ci{art}:   it 
admits  a special $J_2=0$ limit when the $S^5$ part of the solution is 
left (or right) moving. 
Here the string is wrapping a circle of $S^5$ which is not the maximal 
radius one. Explicitly (cf. \rf{solold})\footnote{One may wonder
whether other such solutions exist. One can show that a similar
solution of the form $\Y_{0}=\cosh \rho_{0} \;e^{i\kappa \tau}$,
$\Y_{1}=\sinh \rho_{0} \;e^{i k( \tau-\sigma)}$,
$\X_1=e^{iw\tau+im\sigma}$  does not exist. Also, a solution in
$AdS_{5}$ of the form $\Y_{0}=\cosh \rho_{0} e^{i\kappa \tau}$,
$\Y_{1}=\sqrt{2}\sinh \rho_{0} \;e^{i w \tau+ik\sigma}$,
$\Y_{2}=\sqrt{2}\sinh \rho_{0} \;e^{i m( \tau-\sigma)}$ does not
exist.}
 \bea
  \Y_{0}=r_{0} \; e^{i\kappa \tau}\
 , \quad \Y_{1}=r_{1} \;e^{i\w \tau+ik\sigma}\ ,
  \quad \X_1=\cos \psi_{0}\ e^{im(\tau-\sigma)}\ , \quad \X_2=\sin
  \psi_{0}
 \label{sol}
 \eea
where the coordinates $\rho_{0}$,$\psi_{0}$ specifying the position of the circular string 
 are constant.
 The only non-zero elements of the rotation generators 
 are $S_{50}=E$, $S_{12}=S$, $J_{12}=J$,
 and now $ J=\sqrt{\lambda}m \cos^2 \psi_{0}$.
Again we have 
$
\w^2=k^2+\kappa^2$
and $
2 \kappa \mathcal{E}-\kappa^2=2 \mathcal{S}\sqrt{k^2+\kappa^2}+2
m\mathcal{J}$, $ k\mathcal{S}=m \mathcal{J}$  or 
\begin{equation}
2 \kappa \mathcal{E}-\kappa^2=2 \mathcal{S}\sqrt{k^2+\kappa^2}+2 k 
\mathcal{S} \ .  
\end{equation}
A useful relation following from $ \r_0^2 - \r_1^2 =1 $ is
$
\frac{\mathcal{E}}{\kappa}-\frac{\mathcal{S}}{\sqrt{k^2+\kappa^2}}=1
$
and the non-trivial  solutions for $\k$ are 
\begin{equation}
 \quad \kappa_{\pm}= 2^{-1/2} \sqrt{4 k \mathcal{S}-k^2\pm
k^{3/2}\sqrt{k+8\mathcal{S}}} \ . 
\end{equation}
 Note that $\cos^2 \psi_{0}=\frac{k
\mathcal{S}}{m^2}$. Therefore,  a large $\mathcal{S}$ limit with
$k,m$ held fixed is not well defined. Instead,  a useful limit to
consider is large $\mathcal{S}$ and large $m$ with
$m/\mathcal{S}$=fixed, e.g.,  equal to 1.
 In this limit the string is located near
$\psi_{0}\rightarrow \frac{\pi}{2}$ and $\rho_{0}\rightarrow
\infty$.
The solution $\kappa_{+}$ has a  regular expansion at
small $\mathcal{S}$, which is the flat space limit. 
For the physical $\kappa_{+}$ solution, the energy
$\mathcal{E}=\mathcal{E}(\mathcal{S},k)$ becomes
\begin{equation}
\mathcal{E}=\frac{1}{2}\sqrt{\frac{-k+4\mathcal{S}+\sqrt{k(k+8\mathcal{S})}}{k+4\mathcal{S}+
\sqrt{k(k+8\mathcal{S})}}}\bigg(2\mathcal{S}+\sqrt{2k(k+4\mathcal{S}+\sqrt{k(k+8\mathcal{S})})}\bigg)
\end{equation}
Its large $\mathcal{S}$ expansion is
\begin{equation}
\mathcal{E}=\mathcal{S}+\sqrt{2k}\sqrt{\mathcal{S}}+\frac{k}{4}-\frac{
k^{3/2}}{8\sqrt{2}}\frac{1}{\sqrt{\mathcal{S}}}+\frac{k^2}{32
\mathcal{S}}+O(1/\mathcal{S}^{3/2})
\end{equation}
Let us now consider two ``slow'' limits with small $\mathcal{S}\rightarrow 0$.
The first limit is  $\mathcal{S}\rightarrow 0$ and
$m\rightarrow 0$ or $\mathcal{J}\rightarrow 0$ with $k$ finite. In
this case the string shrinks to a point in both $AdS_{5}$ and
$S^{5}$  and we get the usual flat-space scaling 
\begin{equation} \la{op}
{E}=2\sqrt{k \sql 
{S}}+\frac{{S}^{3/2} }{\l \sqrt{k}}-\frac{5{S}^{5/2}}{4\l^2 k^{3/2}}
+... \ . 
\end{equation}
Another limit is $\mathcal{S}\rightarrow 0$ and
$k\rightarrow \infty$ with $m,\mathcal{J}$ kept finite. Now the string
shrinks to a point  in $AdS_5$ but it remains macroscopic in $S^{5}$.
The energy in this limit is the same as in \rf{op}.
The same result \rf{op} 
is found also in the  special case when   $S=J$, i.e. when  $m=k$.
Thus in contrast with
 a similar limit in the case of the previous $(S,J)$ 
solution, here  we obtain the flat-space behaviour of the energy 
instead of the $\sqrt{\lambda} S $ behaviour.

%We discuss the stability of this solution in Appendix B.

This solution is stable  for sufficiently large $\S$.
One  can compare its  energy 
$\mathcal{E}=\mathcal{E}(k,\mathcal{S})$ with the energy of the
rational solution in the $SL(2)$ sector $\mathcal{E}=\mathcal{E}(k,m,\mathcal{S})$ 
reviewed in the previous subsection.
Numerical analysis shows that 
 the energy of this new  solution 
 has less energy than of the old one. 
 Since the later is known to be stable for large
$\mathcal{S}$, we conclude that the new solution presented here
should also be stable for large $\mathcal{S}$.
This is confirmed by direct analysis of its fluctuation spectrum 
which  follows the discussion in \ci{art}.

\bigskip
\bigskip

\section*{Acknowledgments }
%%%%%%%%%%%%%%%%%%%%%%%%%%%

We are  grateful to   K. Zarembo for useful   comments 
and discussions. We also thank G. Arutyunov and S. Frolov 
for helpful  remarks. 
 The work of A.T. and A.A.T. 
  was supported  in part by the DOE grant DE-FG02-91ER40690.  A.A.T. also
acknowledges the support of PPARC, 
 the INTAS grant  03-51-6346 and the RS Wolfson award.

\renewcommand{\theequation}{A.\arabic{equation}}
 \setcounter{equation}{0}
\setcounter{section}{1} \setcounter{subsection}{0}
 \section*{Appendix: 1-loop correction to  the energy of $(S,J)$ 
solution  in the slow-string limit}

Below we shall consider the  case 
of $m=J$; the discussion  for the general case of 
$m\gg w$ is
similar.

The 1-loop correction to the energy $E_1$ was found in \cite{ptt}.
It can be written as the sum of the contributions of the 
 zero   and non-zero   modes
 \be
 E_1 = E_1^{(0)}  + \bar E_1  \ , \ \ \ \ \ \ \ \ \ 
 E_1^{(0)}=
 { 1 \ov 2\k} ( 4 \nu  + 2\k  +  \w_0  - 8 \td w_0 ) \ , \la{zero}
 \ee
 \bea \la{nonz}
 \bar E_1=  &&{ 1 \ov \k}\sum_{n=1}^\infty
 \bigg( 4 \sqrt{n^2 + \nu^2}+ 2 \sqrt{n^2 + \k^2}
 + \ha  \sum^{4}_{I=1} sign(C_{I,B}^{(n)}) \om_{I,n} \nn\\
 &&\ \ \  - \ 4 \big[ \sqrt{  ( n + c)^2 + a^2}
   + \sqrt{ (n - c)^2 + a^2 }\ \big] \bigg)\ ,  \eea
where $\nu^2 = w^2 - m^2$  and  and for $m=J$ 
\begin{equation}
\omega_{0}=2\sqrt{\kappa^2+J^2(1+r_{1}^2)}, \quad
\tilde{\omega}_{0}=\sqrt{c^2+a^2}\ , \quad
r_{1}^2=\frac{J}{\sqrt{\lambda}\sqrt{J^2+\kappa^2}} \ ,
\end{equation}
\begin{equation}
a^2=\frac{1}{2}\bigg(\kappa^2+\frac{J^2}{\lambda}-J^2\bigg), \quad
c=\frac{1}{2}\kappa\bigg[1+\frac{2J^2(1+r_{1}^2)}{\kappa^2-
\frac{J^2}{\lambda}+J^2}\bigg]\sqrt{\frac{\kappa^2-\frac{J^2}{\lambda}+J^2-2J^2
r_{1}^2}{2(J^2+\kappa^2)}} \ .
\end{equation}
The sign functions are
 \begin{equation}
C_{p}^{B}=\frac{1}{2m_{11}(\w_{p,0})\w_{p,0} \prod_{q\neq
p}(\w_{p,0}^{2}-\w_{q,0}^{2})},\quad
C_{I,B}^{(n)}=\frac{1}{m_{11}(\w_{I,n})\prod _{J\neq
I}(\w_{I,n}-\w_{J,n})}\ ,  \label{signs}
\end{equation}
where $\omega_{I,n}$ are the bosonic frequencies for $n\neq 0$,
$\omega_{p,0}$ are bosonic frequencies for $n=0$, and the relevant
part of the minor $m_{11}$ for computing the signs of
$C_{I,B}^{(n)},C_{p}^{B}$ is $m_{11}\sim(\omega^2-n^2)$.
As  in the
$SU(2)$ case discussed in section 2.2 we
may  expand $E_{1}$ at large $\lambda$ for fixed $J$ and then take $J$ large. 
Again we can do
this expansion inside the sum over $n$. 
Expanding the zero-mode
part we get (omitting the imaginary part) 
\begin{equation}
E_1^{(0)}=1-\sqrt{2}-\frac{5}{4\sqrt{\lambda}}+\frac{27}{32\sqrt{2}\lambda}+...
\end{equation}
The  non-zero mode  bosonic frequencies
from the quartic characteristic equation have the following large
$\lambda$ expansions
\begin{eqnarray}
&&\omega_{I=1,2;n}=-\sqrt{2}J\mp \sqrt{2J^2-2nJ+n^2}\nonumber\\
&&\ \ \ \
-\frac{J}{\sqrt{\lambda}}\frac{6\sqrt{2}J^2+2\sqrt{2}(n-3J)\mp
3(n-2J)\sqrt{2J^2-2Jn+n^2}}{4\sqrt{2}J^2+2(n-2J)(\sqrt{2}n\mp
\sqrt{2J^2-2Jn+n^2})}+O({1\ov \lambda})
\end{eqnarray}
\begin{eqnarray}
&&\omega_{I=3,4;n}=\sqrt{2}J\mp \sqrt{2J^2+2nJ+n^2}\nonumber\\
&&\ \ \ \
+\frac{J}{\sqrt{\lambda}}\frac{6\sqrt{2}J^2+2\sqrt{2}(n+3J)\mp
3(n+2J)\sqrt{2J^2+2Jn+n^2}}{4\sqrt{2}J^2+2(n+2J)(\sqrt{2}n\mp
\sqrt{2J^2+2Jn+n^2})}+O({1\ov \lambda})
\end{eqnarray}
We see
that in the large $\lambda$ limit these frequencies are  real,
so the only unstable modes with $n\leq J$ come from
$S^5$  fluctuations. 
One way to  obtain the real
1-loop correction to energy is to omit the unstable modes, i.e. to 
 take the sum over $n$ starting with 
$n=J$. The sign functions can also be computed in the large
$\lambda$ limit and are found to be
$sign(C_{1,B}^{(n)})=sign(C_{3,B}^{(n)})=-1$ and
$sign(C_{2,B}^{(n)})=sign(C_{4,B}^{(n)})=+1.$ Then we get 
\begin{equation}
\bar E_1=\sum_{n=1}^{\infty}S_{n}(J,\lambda), \quad\quad
S_{n}=B_{0}+\frac{B_{1}}{\sqrt{\lambda}}+\frac{B_{2}}{\lambda}+...
\ ,
\end{equation}
where
\begin{eqnarray}
B_{0}&=&\frac{1}{J}\bigg[4\sqrt{n^2-J^2}+2\sqrt{n^2+J^2}+\sqrt{n^2-2nJ+2J^2}+\sqrt{n^2+2Jn+2J^2}\nonumber\\
&-& 2\sqrt{2}|J-\sqrt{2}n|-2\sqrt{2}|J+\sqrt{2}n|\bigg] \ .
\end{eqnarray}
$B_{1}$, $B_{2}$ have complicated form which we will not write down,
but we used them to evaluate the series numerically and plot $\bar E_1 $   against
$J$. Taking the sums over $n$ from $J$ to $N=10^5$ we plotted
$B_{0}$, $B_{1}$ and $B_{2}$ for  $J=10^2,...,10^4$. As
in the $SU(2)$ case, we  found linear dependence with $J$. 
Combining together the classical energy and the 1-loop correction we got\foot{
Note that again  the  zero-mode part does not contribute in the large
$J$ limit.}
\begin{equation}
E=J\bigg[\sqrt{\lambda}\bigg(1+(\sqrt{2}-\frac{\hbar}{4})\frac{1}{\sqrt{\lambda}}
\bigg)+
\frac{1}{\sqrt{\lambda}}\bigg(\frac{1}{4}-\frac{\hbar}{1.75}\bigg)+
\frac{1}{\lambda}\bigg(7.5-\frac{\hbar}{8\sqrt{2}}\bigg)+O(\lambda^{-3/2})\bigg]\ . 
\end{equation}
 Here as in \rf{qenergy} we introduced $\hbar$ 
 to indicate the 1-loop contributions.

\end{document}